# Cross-coupled photon avalanche drives room-temperature intrinsic optical bistability

Eva Conour, Jacob Wojcik, Jory Sanders, Abigale Puccini, Trey Stephens-Cherry, and Artiom Skripka*

Department of Chemistry, Oregon State University, Corvallis, OR 97331, USA

*artiom.skripka@oregonstate.edu

## Abstract

Optically bistable materials promise technological advances in optical information processing[1,2], memristive devices[3], and multiparametric sensing[4,5]. However, optical bistability is predominantly observed under cryogenic conditions[6] or is governed by subtle thermo-optical nonlinearities[7,8]. Here, we report on $Ho^{3+}$-activated nanocrystals that exhibit cavity-free intrinsic optical bistability (IOB) at ambient conditions. Non-thermal IOB stems from a cross-coupled photon avalanche (XPA) mechanism enabled by a low-phonon-energy double-perovskite host[9], which reduces nonradiative losses and facilitates excited-state population via XPA, even when the pump intensity is below threshold. Time-resolved luminescence measurements support the proposed XPA, as evidenced by a double-rise profile consistent with two reservoir states interlocked in energy looping. IOB with high-contrast photoswitching (up to 46 dB) is observed at room temperature across multiple visible emission lines, enabling wavelength-division multiplexing and compatibility with common detectors. IOB is preserved across nano- to microscale crystals and different double-perovskite compositions, broadly enabling information photonics and advanced sensing.

Exceptionally nonlinear luminescence in photon-avalanche nanoparticles (ANPs) has enabled new capabilities, including sub-100 nm optical imaging[10–12], sensing of piconewton forces[13], and all-optical computing[14,15]. The photon avalanche (PA) response, characterized by a sharp intensity increase above an irradiance threshold, arises from autocatalytic excited-state population between lanthanide ($Ln^{3+}$) ions undergoing repeated excited-state absorption (ESA) and cross-relaxation (CR) events[16]. When the positive feedback of PA can outcompete nonradiative decay within a finite range of pump intensities below the threshold, two distinct luminescence outputs persist under identical pump conditions, with a particular output defined by excitation history[17]. As a result, these optically bistable materials exhibit pump-dependent luminescence hysteresis, which enables optical memory and logic-level operations, as well as multiparametric sensing of physical and chemical cues. However, cavity-free intrinsic optical bistability (IOB) is rare and often thermally driven[18], limiting control over optical-state switching and material applications.

Non-thermal IOB was reported for $KPb_2Cl_5$:$Nd^{3+}$ ANPs operated at 77 K, where the low-phonon-energy $KPb_2Cl_5$ matrix strongly suppressed multiphonon relaxation (MPR) and activated energy transfer between otherwise quenched low-lying $Nd^{3+}$ excited states[17]. Such optically bistable ANPs performed transistor-like photoswitching with bit-level optical memory. However, the $KPb_2Cl_5$ host can incorporate only moderate concentrations of $Nd^{3+}$ ions and is inhospitable to conventional PA systems based on heavier lanthanides ($Tm^{3+}$, $Ho^{3+}$, and $Pr^{3+}/Yb^{3+}$)[19,20]. The mismatch in ionic radii and charge makes substituting $Pb^{2+}$ with $Ln^{3+}$ ions unfavorable; therefore, critical activator concentrations cannot be reached for sufficiently high CR rates and, subsequently, PA. Furthermore, observing IOB in $KPb_2Cl_5:Nd^{3+}$ requires cryogenic cooling, adding cost and operational complexity. As a result, achieving IOB under ambient conditions in ANPs remains a materials-design challenge, requiring lanthanide-host combinations that favorably modulate optoelectronic and phononic processes to sustain excited-state population.

Here, we demonstrate IOB in $Ho^{3+}$-activated heavy-halide ANPs based on the $Cs_2NaInCl_6$ matrix at ambient conditions. We observe binary-like photoswitching between weakly and brightly luminescent $Ho^{3+}$ states under a 976 nm pump, as well as pump-history-dependent luminescence hysteresis. This IOB property arises from suppressed MPR in a low-phonon-energy host and from unique nonlinear positive feedback in a cross-coupled photon avalanche (XPA) encoded in the $Ho^{3+}$ energy-level structure. Unlike other PA mechanisms, XPA depends on two interlocked reservoir states that are cross-populated by non-degenerate CR and participate in a two-step ESA (Fig. 1a). As a result, XPA engages high-lying $Ho^{3+}$ energy states in energy looping and provides a bistable response across multiple visible emissions. We observed IOB of $Ho^{3+}$ only in the low-phonon-energy chloride host, and it is retained across different particle length scales and double-perovskite compositions.

Double perovskites with the canonical $A_2B(I)B(III)X_6$ elpasolite structure (A – alkali metal, B(I) – monovalent cation, B(III) – trivalent cation, X – halide) provide a compositionally diverse, low-phonon-energy (<300 $cm^{-1}$; ref. 21) platform, where the B(III) site can isovalently host $Ln^{3+}$ ions (Fig. 1b). These materials are less toxic alternatives to lead halide perovskites and are stable under ambient conditions[22]. To synthesize $Cs_2NaInCl_6$ nanocrystals activated with different $Ho^{3+}$ concentrations, we used a modified hot-injection method (Supplementary Section 1), which yielded cubic-phase nanocrystals with an average size of 56 nm (Fig. 1c, Fig. S1, S2). Upon 976 nm excitation, which is non-resonant with $Ho^{3+}$ ground-state absorption (GSA) but resonant with several ESA transitions, the $Cs_2NaInCl_6$: 40% $Ho^{3+}$ nanocrystals exhibited visible luminescence in the 450-750 nm range (Fig. 1d). In addition to the commonly observed emissions at 490 nm ($^5F_3 \rightarrow {}^5I_8$), 540 nm ($^5F_4/{}^5S_2 \rightarrow {}^5I_8$), and 660 nm ($^5F_5 \rightarrow {}^5I_8$), we also detected an unconventional orange emission at 590 nm. We ascribe this emission to a highly energetic $^5G_4$ state (at 25715 $cm^{-1}$; ref. 23) relaxing to the $^5I_6$ excited state of $Ho^{3+}$ (Fig. 1e). This $^5G_4 \rightarrow {}^5I_6$ transition was recently reported in an analogous low-phonon-energy $Cs_2NaYCl_6$ host[24], and is distinct from the 578 nm ($^5F_1 \rightarrow {}^5I_7$) transition observed in fluoride hosts.

We then investigated whether $Cs_2NaInCl_6:Ho^{3+}$ nanocrystals exhibit PA by performing power-dependent steady-state measurements at room temperature. At the lower 976 nm pump intensity, only weak 660 nm emission was observed (Fig. 1d); however, the upconversion intensity increased abruptly above the 200 $kW \cdot cm^{-2}$ threshold, with all identified $Ho^{3+}$ transitions detectable. On average, $Cs_2NaInCl_6:Ho^{3+}$ nanocrystals showed nearly a 10-fold increase in luminescence intensity above the threshold (Fig. 1f), analogous to the binary-like photoswitching of cryogenically cooled $KPb_2Cl_5:Nd^{3+}$ ANPs. This highly nonlinear behavior of $Cs_2NaInCl_6:Ho^{3+}$ points to a PA process governing the excited-state population. Strikingly, a reverse power scan revealed that $Cs_2NaInCl_6:Ho^{3+}$ ANPs continued to photoluminesce even when the pump intensity dropped below the initial switch-on threshold, down to 150 kW·cm$^{-2}$. As a result, the complete power-dependent measurement exhibited luminescence hysteresis. The $^5G_4 \rightarrow ^5I_6$ transition (590 nm) showed the same hysteresis profile (Fig. 1f), indicating that IOB in $Cs_2NaInCl_6:Ho^{3+}$ sustains population of high-lying excited states in its photoactivation mechanism. $Cs_2NaInCl_6:Ho^{3+}$ ANPs exhibited activator-concentration-dependent nonlinear behavior, with IOB appearing when the nominal $Ho^{3+}$ concentration reached 40% (Fig. S3, Table S1). Such high B(III) site-substitution levels are necessary to counteract the large interionic distances in double perovskites[25], thereby facilitating energy transfer via CR once critical activator concentrations are reached.

To confirm that the low phonon energies of the heavy halide hosts are key to observing IOB at room temperature, we performed control experiments on $NaYF_4:Ho^{3+}$ and $LiYF_4:Ho^{3+}$ nanocrystals (Supplementary Section 4). Because of their higher phonon cutoff frequencies, the upconversion signal is dominated by the 660 nm emission ($^5F_5 \rightarrow ^5I_8$), and neither composition showed luminescence at 590 nm even at a 976 nm pump intensity of 1 MW·cm$^{-2}$ (Fig. S9). The $NaYF_4:Ho^{3+}$ nanocrystals showed strongly nonlinear PA upconversion (nonlinearity = 11) without hysteresis, whereas $LiYF_4:Ho^{3+}$ exhibited only minor energy looping (nonlinearity = 5.5) (Fig. S10). Based on the comparison between chloride and fluoride hosts, we conclude that the low phonon cutoff frequencies of chlorides favor multicolor upconversion and IOB by $Ho^{3+}$ activator ions.

Drawing on the steady-state data and IOB behavior observed from the highly energetic $^5G_4$ state, we hypothesized that two ESA steps, resonant with a 976 nm pump, are involved in energy looping, specifically ESA1 ($^5I_7 \rightarrow ^5F_5$) and ESA2 ($^5F_5 \rightarrow ^5G_4$). Therefore, the corresponding $^5I_7$ and $^5F_5$ reservoir states should be populated concurrently via a non-degenerate CR, forming a cross-coupled PA (XPA) energy loop. Because ESA1 terminates and ESA2 originates from the $^5F_5$ reservoir state, we conducted power-dependent time-resolved photoluminescence measurements at 660 nm to elucidate the proposed XPA mechanism. At pump intensities corresponding to the pre-avalanche regime, we observed ordinary single-rise and decay profiles (Fig. 2a). However, as the pump intensity reached the avalanche regime, a two-step rise profile emerged, and the decay time of the $^5F_5$ state decreased from 5.9 ms at an average irradiance of 75 kW·cm$^{-2}$ to 3.2 ms at 145 kW·cm$^{-2}$, indicating an additional depopulation pathway. Notably, the second rise becomes steeper and occurs earlier as irradiance increases (Fig. 2a, Fig. S4), consistent with higher photon fluxes establishing a critical $^5F_5$ population and activating XPA (ESA1+ESA2+CRs) sooner. This also agrees with a defining PA criterion: the $Cs_2NaInCl_6:Ho^{3+}$ system shows slow rise times near the threshold and faster rise times at higher pump intensities, as reflected in the dynamics of the second rise profile. Further analysis of the time delay between the *dI/dt* extrema (Fig. S5) of the first and second rises shows a superlinear decrease ($-\alpha = -1.43$, slope in the log-log plot of Fig. 2b) with increasing irradiance. We attribute this to a cross-coupling mechanism in which both reservoir states $^5I_7$ and $^5F_5$ are interlocked via two-step $^5I_7 \rightarrow ^5F_5 \rightarrow ^5G_4$ ESA and populated concurrently via CR, defining distinct features of XPA.

To support our proposed XPA mechanism, we conducted differential rate-equation modeling of $Ho^{3+}$ photoactivation dynamics under a 976 nm pump (Supplementary Section 5). Simulated upconversion spectra supported the assignment of the 590 nm emission to the $^5G_4 \rightarrow ^5I_6$ transition (Fig. S12, S13). Simulations of the $NaYF_4:Ho^{3+}$ system did not predict this emission and indicated that MPR is prevalent in most excited states. The power-dependent simulations also reproduced the bistable behavior of $Ho^{3+}$ and the double-rise profile (Fig. S14) and allowed us to identify a set of non-degenerate CR pathways, CR1: [$^5F_3$, $^5I_8$] → [$^5F_5$, $^5I_7$], CR2: [$^5F_3$, $^5I_8$] → [$^5I_7$, $^5F_5$], CR3: [$^5G_4$, $^5I_8$] → [$^5F_3$, $^5I_7$], and CR4: [$^5F_4$/$^5S_2$, $^5I_8$] → [$^5I_4$, $^5I_7$] that may contribute to energy looping (Fig. 2c). CR1 and CR2 directly cross-populate the $^5I_7$ and $^5F_5$ PA reservoir states, while CR3 helps populate $^5F_3$, which facilitates CR1 and CR2. Experimentally, we observed that increasing the $Ho^{3+}$ concentration decreased the relative contribution of the 590 nm band in the upconversion spectrum (Fig. S6) and shortened the $^5G_4$ state lifetime; this is consistent with concentration quenching and supports feeding of the $^5F_3$ excited state via CR3 (Fig. S7). Each CR, including CR4, populates the first reservoir state ($^5I_7$) and one of the higher excited states, collectively outcompeting MPR-induced quenching. In addition, CR-knockout simulations show that $Ho^{3+}$ IOB in low-phonon-energy hosts vanishes when CR1 and CR2 are deactivated (Fig. S15, S17). XPA therefore persists as long as a suitable CR pathway can concurrently populate both $^5I_7$ and $^5F_5$ reservoirs. Unlike parallel photon avalanche in $Ho^{3+}$, where two independent reservoir states sustain individual loops that populate distinct emissive levels for tunable chromaticity[26], XPA forms an energy loop through CRs that terminate at two reservoir states that are interlocked via a two-step ESA (Fig. 2c). Therefore, in low-phonon-energy hosts, XPA is protected from MPR even at room temperature. Multiple CR pathways continuously return population to the two reservoirs, and an excited ion must cascade through many intermediate levels to relax thermally to the ground state. This distinguishes XPA-derived IOB from the conventional PA-driven IOB of $KPb_2Cl_5$:$Nd^{3+}$ ANPs, which persists only at 77 K because a single reservoir state ($^4I_{11/2}$) is close to the ground state ($^4I_{9/2}$) and requires cooling to suppress MPR further.

Having unveiled the mechanism of $Ho^{3+}$ avalanche upconversion and IOB, we explored whether XPA extends to other double perovskite matrices. We synthesized $Cs_2NaB(III)Cl_6$:$Ho^{3+}$ (B(III) = In, Y, Lu) microcrystals by dissolution-recrystallization and characterized their optical properties (Supplementary Section 6). Consistent with $Cs_2NaInCl_6$:$Ho^{3+}$ ANPs, all $Ho^{3+}$-activated microcrystalline powders exhibited upconversion emission under 976 nm excitation (Fig. 3a) with power-dependent hysteresis (Fig. 3b, Fig. S21-23, Supplementary Video 1). Time-resolved analysis also confirmed a double-rise profile of the $^5F_5$ excited-state population in $Cs_2NaInCl_6$:$Ho^{3+}$ microcrystals, with the time delay between the first and second rises decreasing superlinearly as pump intensity increased (Fig. S24-26). As a result, XPA and room-temperature IOB are maintained in $Ho^{3+}$-activated double perovskites regardless of their size and composition.

To evaluate the performance of optically bistable XPA materials, we analyzed measurements taken at multiple spots across films of $Cs_2NaInCl_6$:$Ho^{3+}$ ANPs and microcrystalline powders. On average, $Cs_2NaInCl_6$:$Ho^{3+}$ ANPs switched on at a pump intensity of 206 $kW\cdot cm^{-2}$ (equivalent to 1.1 mW), with a hysteresis width of 46 $kW\cdot cm^{-2}$. The $Cs_2NaInCl_6$:$Ho^{3+}$ microcrystalline sample exhibited a switch-on pump intensity of 125 $kW\cdot cm^{-2}$ and a hysteresis width of 40 $kW\cdot cm^{-2}$. Compared with the previous study of $KPb_2Cl_5$:$Nd^{3+}$ nanocrystals, $Ho^{3+}$-activated double perovskites require higher excitation thresholds but exhibit hysteresis over a wider power range and do not require cryogenic cooling (Fig. 3c). These power thresholds still fall within the typical range of 10-600 $kW\cdot cm^{-2}$ pump intensities used to activate conventional ANPs[10,11,27]. As a figure of merit, we calculated the photoswitching contrast for the 660 nm emission (i.e., the signal-to-noise ratio). $Cs_2NaInCl_6$:$Ho^{3+}$ ANPs at room temperature have a switch-on contrast of 23.7±4.3 dB, which is 10 dB (an order of magnitude) greater than $KPb_2Cl_5:Nd^{3+}$ ANPs at 77 K (Fig. 3d). The best-performing double perovskite, $Cs_2NaLuCl_6:Ho^{3+}$, reached a switching contrast of 46.3±7.1 dB (Fig. 3d). To the best of our knowledge, this is the highest switching contrast reported for $Ln^{3+}$-activated materials and rivals that of cavity-based optically bistable devices[28,29].

In summary, we developed (nano)materials that exhibit intrinsic optical bistability at room temperature, which we attribute to the low phonon energies of the heavy-halide host and a unique cross-coupled photon avalanche (XPA) mechanism encoded in $Ho^{3+}$ ions. The resulting compounds show near-binary photoswitching with contrast up to 46.3 dB, wide luminescence hysteresis, and IOB across particle sizes and compositions, indicating that bistability is intrinsic to the $Ho^{3+}$-activated double perovskites. Accessible room-temperature IOB in the visible spectral range opens possibilities for wavelength-multiplexed all-optical memory and logic, optical thresholding, sensing, and cryptographic primitives. Key open questions concern how far host and surface engineering, machine-learning-aided optimization, and photonic-structure coupling can push the dynamics that govern switching speed and thresholds. Overall, XPA recasts intrinsic optical bistability as a materials-design target for a tunable family of bistable emitters.

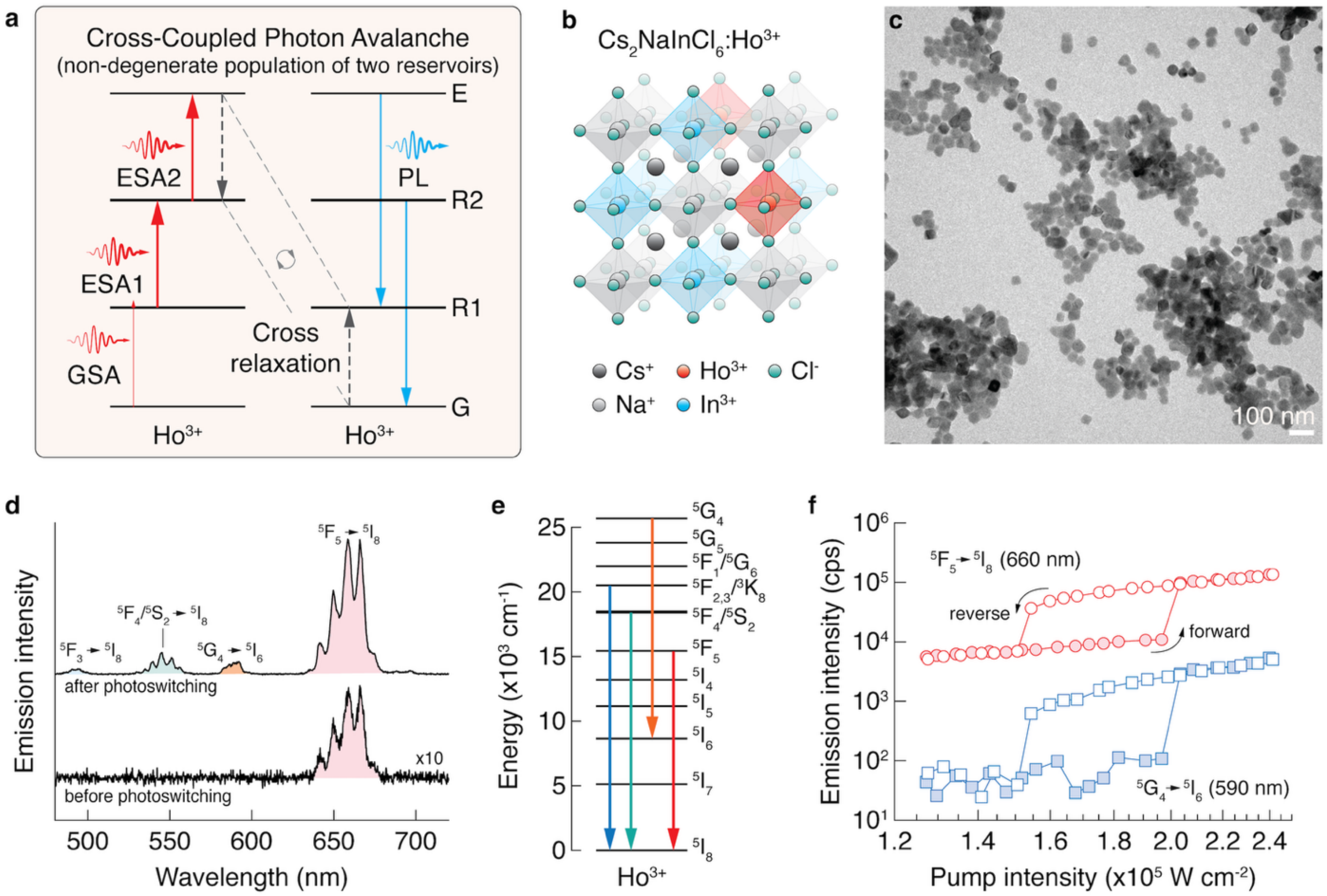

**Figure 1. The concept of XPA and observation of IOB in $Ho^{3+}$-activated double perovskites.** **a**, Generalized scheme of XPA: two interlocked reservoir states (R1 and R2) are cross-coupled via CR and participate in two-step ESA processes. GSA – ground-state absorption, ESA – excited-state absorption, PL – photoluminescence, G – ground, R – reservoir, E – excited states. **b**, Corner-sharing octahedral structure of the double perovskite matrix. **c**, Transmission electron microscopy (TEM) image of $Cs_2NaInCl_6$: 40% $Ho^{3+}$ ANPs. **d**, Photoluminescence spectrum of $Cs_2NaInCl_6:Ho^{3+}$ ANPs before and after the switch-on threshold. Signal intensity before photoswitching is multiplied by 10 for visibility. **e**, Assigned radiative transitions of $Cs_2NaInCl_6:Ho^{3+}$ ANPs. **f**, Pump power dependence of $Cs_2NaInCl_6:Ho^{3+}$ ANPs emission at 590 nm ($^5G_4 \rightarrow {}^5I_6$; blue squares) and 660 nm ($^5F_5 \rightarrow {}^5I_8$; red circles) at room temperature. Filled data points represent PL intensity acquired with increasing pump intensity (forward scan), and open points with decreasing pump intensity (reverse scan).

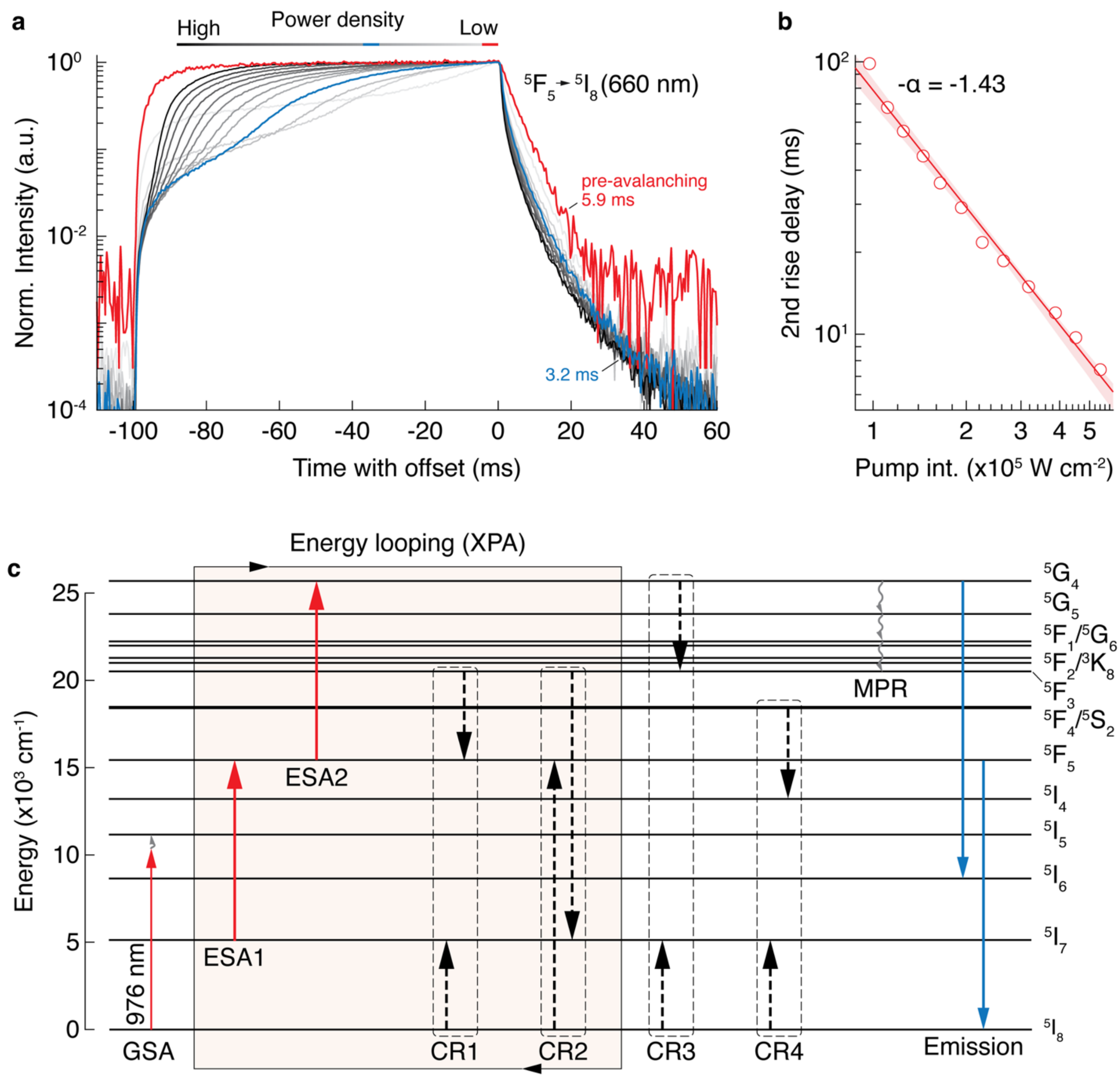

**Figure 2. Photoactivation dynamics of XPA.** **a**, Time-resolved photoluminescence of $Cs_2NaInCl_6:Ho^{3+}$ ANPs at different 976 nm pump intensities. The 976 nm pump was pulsed at 4 Hz with a 50% duty cycle. Representative traces of the pre-avalanching (red) and avalanching (blue) regimes are highlighted to emphasize a double-rise profile and shortened lifetimes. Traces are time-offset to distinguish rise and decay profiles relative to the falling edge of the excitation pulse. Mean lifetimes were calculated as the area under the curve of the decay profile. **b**, Log-log plot of the 2$^{nd}$ rise delay vs pump intensity. Experimental points (open circles) are fitted with a linear regression (red line), and the shaded area indicates the 95% confidence interval. The slope of the fit is $\alpha$. **c**, XPA mechanism of $Ho^{3+}$ showing GSA, ESAs, CRs, MPRs, and emissions that contribute to energy looping and IOB. Non-degenerate CR1 and CR2 terminate directly at the $^5F_5$ and $^5I_7$ reservoir states, which close the loop via ESA1 and ESA2 transitions.

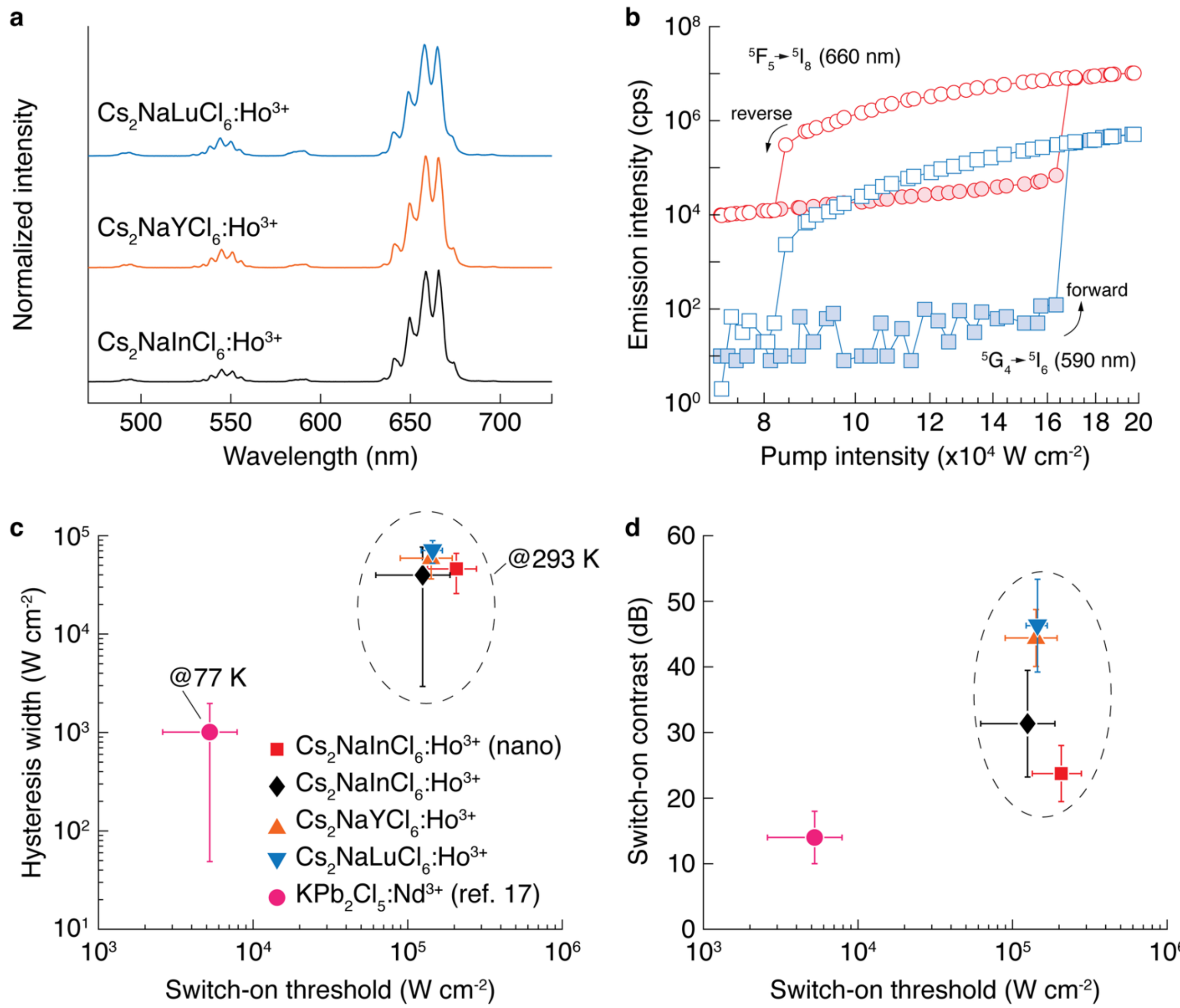

**Figure 3. Extension and performance of room-temperature IOB.** **a**, PA upconversion spectra of $Cs_2NaB(III)Cl_6:Ho^{3+}$ (B(III) = In, Y, Lu) microcrystals under 976 nm excitation. **b**, Pump power-dependent emission at 590 nm ($^5G_4 \rightarrow {}^5I_6$; blue squares) and 660 nm ($^5F_5 \rightarrow {}^5I_8$; red circles) of $Cs_2NaLuCl_6:Ho^{3+}$ microcrystals at room temperature. Filled data points represent the PL intensity measured at increasing pump intensity (forward scan), and open ones at decreasing pump intensity (reverse scan). **c**, Comparison of hysteresis width, and **d**, switch-on contrast vs switch-on threshold, among different double perovskite materials studied at room temperature and $KPb_2Cl_5:Nd^{3+}$ ANPs at 77 K (ref. 17). Figure of merit values are shown as mean values ± one standard deviation (n = 32 for $Cs_2NaInCl_6:Ho^{3+}$ ANPs, n = 13 for $Cs_2NaInCl_6:Ho^{3+}$ and $Cs_2NaLuCl_6:Ho^{3+}$ microcrystals, n = 10 for $Cs_2NaYCl_6:Ho^{3+}$ microcrystals).

## Methods

The Supplementary Information provides all experimental details, including nanocrystal and microcrystal synthesis, structural characterization, optical measurements, and simulations.

## Data and Materials Availability

All data in the main text or the supplementary materials are available from the corresponding author upon reasonable request.

## Acknowledgments

The authors thank Emory Chan and Xiaogang Liu for their valuable and insightful discussions, and Prayash Mohanti and Kyriakos Stylianou for assistance with photoluminescence excitation measurements. A.S. thanks Oregon State University for startup funds. E.C., J.S., and J.W. were in part supported by the URSA Engage program of Oregon State University.

## Author Contributions

E.C. and A.S. designed and conceptualized the research. A.S. directed the study. E.C., T.S.-C., and A.S. developed and performed optical characterization experiments. J.W. and A.S. performed numerical simulations and data analysis. E.C. and A.P. synthesized and characterized nanocrystals. J.W. and J.S. synthesized and characterized microcrystals. E.C. and A.S. wrote the original manuscript. All authors discussed the results and revised the manuscript.

## Competing Interests

The authors declare that they have no competing interests.

## Additional Information

Supplementary Information (materials and methods, Figs. S1 to S26, Tables S1 to S5, supplementary references); Supplementary Data 1; Supplementary Video 1.

## Supplementary Information

# Cross-coupled photon avalanche drives room-temperature intrinsic optical bistability

Eva Conour, Jacob Wojcik, Jory Sanders, Abigale Puccini, Trey Stephens-Cherry, and Artiom Skripka*

Department of Chemistry, Oregon State University, Corvallis, OR 97331, USA

*artiom.skripka@oregonstate.edu

**Table of Contents**

**Supplementary Section 1. Materials and Methods**

## 1.1 Materials

Cesium carbonate ($Cs_2CO_3$, 99%), sodium acetate, anhydrous ($CH_3COONa$, ≥99.0%), indium (III) acetate hydrate ($In(CH_3COO)_3{\cdot}xH_2O$, 99.99%), yttrium (III) acetate hydrate ($Y(CH_3COO)_3{\cdot}xH_2O$, 99.99%), yttrium (III) chloride hexahydrate ($YCl_3{\cdot}6H_2O$, 99.99%), holmium (III) acetate hydrate ($Ho(CH_3COO)_3{\cdot}xH_2O$, 99.99%), holmium (III) chloride hexahydrate ($HoCl_3{\cdot}6H_2O$, 99.9%), cesium chloride (CsCl, 99.9%), indium (III) chloride tetrahydrate ($InCl_3{\cdot}4H_2O$, 97%), lutetium (III) chloride hexahydrate ($LuCl_3{\cdot}6H_2O$, 99.99%), europium (III) chloride hexahydrate ($EuCl_3{\cdot}6H_2O$, 99.99%), ammonium fluoride ($NH_4F$, 99.9%), myristoyl chloride (97%), oleic acid (OA, 90%), octadecene (ODE, 90%), oleylamine (OM, 70%), trifluoroacetic acid (99%) and toluene (≥99.5%) were purchased from Sigma-Aldrich. Lithium trifluoroacetate (Li-TFA, 97%), yttrium (III) oxide ($Y_2O_3$, 99.999%), holmium (III) oxide ($Ho_2O_3$, 99.99%), and N-octyl-N,N-dimethyl-3-ammonio-1-propanesulfonate (octyl sulfobetaine, 98%) were purchased from Thermo Fisher Scientific. Sodium chloride (NaCl, 99%) was purchased from VWR Chemicals. Sodium oleate (>97.0%) was purchased from TCI. All chemicals were used without any further purification.

## 1.2 Synthesis of $Cs_2NaInCl_6$:$Ho^{3+}$ nanoparticles

$Ho^{3+}$-activated $Cs_2NaInCl_6$ photon avalanching nanoparticles (ANPs) were prepared by hot-injection using a modified colloidal synthesis method.[1] In a typical synthesis of $Cs_2NaInCl_6$: 40% $Ho^{3+}$ ANPs, $Cs_2CO_3$ (0.03584 g), sodium acetate (0.01805 g), indium acetate (0.04379 g), holmium acetate (0.03421 g), OA (1.25 mL), ODE (5 mL), OM (0.33 mL), and octyl sulfobetaine (0.0279 g) were added into a 50 mL three-neck flask and the solution was set to stir (800 rpm) under a $N_2$ flow using a Schlenk line. After a quick $N_2$ purge, the flask was placed under vacuum, and the temperature was increased to 110 °C in 20 °C increments. The solution was maintained under vacuum for 90 minutes, followed by two $N_2$ purges, with 5 minutes under vacuum after each purge. The solution was then placed under nitrogen flow, heated to 240 °C, and held for 15 minutes to equilibrate. 300 μL of myristoyl chloride was quickly injected, and the reaction proceeded for 2 minutes before quenching in a water bath. The contents were then transferred to a 50 mL centrifuge tube and centrifuged at 4700 rpm for 30 minutes. The supernatant was discarded, and the pellet was redispersed in 4 mL toluene, then centrifuged at 4700 rpm for 10 minutes. The supernatant was discarded, and the pellet was redispersed in 4 mL of toluene for a final centrifugation at 1000 rpm for 2 minutes. The resulting supernatant was collected into a glass vial and stored under ambient conditions for characterization. $Cs_2NaInCl_6$ ANPs containing different concentrations of $Ho^{3+}$ ions were synthesized using the same method, with stoichiometrically adjusted amounts of indium acetate and holmium acetate.

## 1.3 Synthesis of $NaYF_4$:$Ho^{3+}$ nanoparticles

$NaYF_4$:$Ho^{3+}$ nanoparticles were synthesized using a previously described synthesis,[2] with some modification. To a dry 50 mL 3-neck round-bottom flask, yttrium chloride (0.8 mmol, 242.7 mg), holmium chloride (0.2 mmol, 75.9 mg), OA (6 mL), and ODE (14 mL) were added. The flask contents were stirred, placed under vacuum, and heated to 100 °C for 1 hour, until the solution became clear. The flask was then filled with $N_2$, and sodium oleate (2.5 mmol, 761.1 mg) and $NH_4F$ (4 mmol, 148.1 mg) were added. The flask was then placed under vacuum and stirred for another 20 min, followed by three $N_2$ flushes. The reaction was heated to 320 °C and allowed to react under $N_2$. After 40 min, the flask was cooled to room temperature with a strong stream of air, and the nanoparticles were isolated by centrifugation (4000 rpm, 4 min) using ethanol (20 mL). The pellet was resuspended in 5 mL of hexane and centrifuged (4000 rpm, 2 min) to separate and discard the insoluble particles (pellet). The nanoparticles were washed twice with hexane:ethanol (1:1 *v*/*v*) and redispersed in 4 mL of hexane for storage.

## 1.4 Synthesis of $LiYF_4$:$Ho^{3+}$ nanoparticles

$LiYF_4:Ho^{3+}$ nanoparticles were prepared via thermal decomposition of metal trifluoroacetate precursors.[3] First, yttrium and holmium trifluoroacetate precursors were prepared from their respective oxides. To a 100 mL 3-neck round-bottom flask, $Y_2O_3$ (1 mmol, 225.81 mg) and $Ho_2O_3$ (0.25 mmol, 94.47 mg) were added together with 5 mL trifluoroacetic acid and 5 mL of distilled water; the mixture was refluxed under vigorous stirring at 80 °C until clear, at which point the temperature was decreased to 60 °C to slowly evaporate the trifluoroacetic acid and water. The precursors were obtained as solids, dried, and used without further purification.

Subsequently, Li-TFA (2.5 mmol, 309.2 mg), 20 mL of OA, and 20 mL of ODE were added to the dried precursors. The mixture was degassed at 120 °C under vacuum with magnetic stirring for 45 min, followed by two $N_2$ flushes. The reaction was then heated to 320 °C and allowed to react under $N_2$. After 60 min, the flask was cooled to room temperature with a strong stream of air, and the nanoparticles were isolated by centrifugation (4000 rpm, 10 min) using ethanol. The nanoparticles were washed twice with hexane:ethanol (1:1 *v*/*v*) and redispersed in hexane for storage.

### 1.5 Synthesis of $Cs_2NaB(III)Cl_6$: 40% $Ho^{3+}$ (B(III)=In, Y, Lu) microcrystals

Adapting a previously reported protocol,[4] CsCl, NaCl, $B(III)Cl_3 \cdot nH_2O$, and $HoCl_3 \cdot 6H_2O$ were added to a 20 mL vial and dissolved in 10 mL of deionized water in the ratio 2:1:0.6:0.4 (total of 5 mmol). The mixture was then sonicated for 20 minutes, and the closed vial was placed in an oil bath heated to 80 °C for two hours. The oil bath temperature was then lowered to 70 °C, the cap was removed, and the microcrystals were left to dry overnight.

### 1.6 Synthesis of $Cs_2NaInCl_6$: 5% $Eu^{3+}$ microcrystals

The synthesis procedure for $Cs_2NaInCl_6$: 5% $Eu^{3+}$ microcrystals was identical to the previously described procedure; the only difference was substituting $EuCl_3 \cdot 6H_2O$ for $HoCl_3 \cdot 6H_2O$.

### 1.7 Characterization

Characterization via X-ray diffraction (XRD) was performed using a Rigaku MiniFlex600 benchtop X-ray diffractometer in the 2θ range 20-60° using Cu Kα radiation. Transmission electron microscopy (TEM) was performed using the FEI Titan 80-200/ChemiSTEM microscope. Photoluminescence excitation (PLE) spectra of $Cs_2NaInCl_6:Eu^{3+}$ microcrystals were acquired using a PTI Quantmaster 8000 series fluorometer (HORIBA Scientific) with a xenon arc lamp, running on FelixGX software. 615 nm emission was monitored while scanning the excitation wavelength across the 350-550 nm range. Concentrations of $Ho^{3+}$ ions in $Cs_2NaInCl_6:Ho^{3+}$ nanocrystals (**Supplementary Table S1**) were determined using inductively coupled plasma optical emission spectroscopy (ICP-OES, Spectro Arcos). Samples were prepared by digesting dry nanocrystals in 3% w/w $HNO_3$. $Ho^{3+}$ and $In^{3+}$ ion calibration standards were purchased from Inorganic Ventures.

### 1.8 Photoluminescence measurements

Photon avalanche emission from $Ho^{3+}$-activated double perovskites was measured in films prepared by drop-casting toluene dispersions of nanocrystals onto glass coverslips (0.13 mm thickness) or by depositing powder samples of microcrystals directly onto the coverslips.

Samples were characterized using a custom-built confocal inverted microscope (**Supplementary Scheme S1**). A 976 nm continuous-wave laser source (TEM-F-975, Changchun New Industries Optoelectronics Tech. Co., Ltd.) was coupled through a 950 nm short-pass dichroic mirror (DMSP950R, Thorlabs) and a 40× 0.95 NA air objective. Objective height (i.e., focus) was controlled with a voice-coil PIFOC focus drive (V-308.753030, Physik Instrumente) operated by a PIMag® motion controller (C-414.13030, Physik Instrumente). Sample positioning was performed using a linear-motor XY microscope stage (X-ADR130B100B-SAE53D12, Zaber Technologies). Photoluminescence was collected by the same objective, with additional spectral filtering by a 950 nm short-pass filter (FF01-950/SP-25, Semrock). Spectral data were recorded using a CCD camera (iDUS 420A-BEX2-DD, Andor, Oxford Instruments) equipped with a spectrograph (Kymera-328I, Andor, Oxford Instruments). Intensities (in counts per second; cps) of $Ho^{3+}$ emission lines at 590 nm and 660 nm were measured using a pair of avalanche photodiodes (SPCM-NIR, Excelitas Technologies) concurrently via a photon-counting module (MultiHarp 150, PicoQuant). 590/10 nm (FBH590-10, Thorlabs) and 650/60 nm (FF01-650/60-25, Semrock) band-pass filters, together with a 604 nm short-pass dichroic mirror (AF604-DI01-25X36, Semrock), were used to spectrally select emissions. The same acquisition module recorded time-resolved photoluminescence, synchronized with an optical chopper (MC2000B with MC1F2, Thorlabs) that modulated the pump laser. For power-dependence measurements, a circularly variable metallic neutral density filter (100FS02DV.2, Newport) was inserted in the laser beam and controlled by motorized rotation. Powers were simultaneously recorded by a Thorlabs power meter (PM101A with S120C, Thorlabs) from a glass coverslip that reflected ~10% of the incoming flux. Average excitation power densities were calculated using measured laser powers and the $1/e^2$ area for the employed excitation wavelength and microscope objective. Instrument control, automation, and data acquisition were performed using the open-source ScopeFoundry platform.

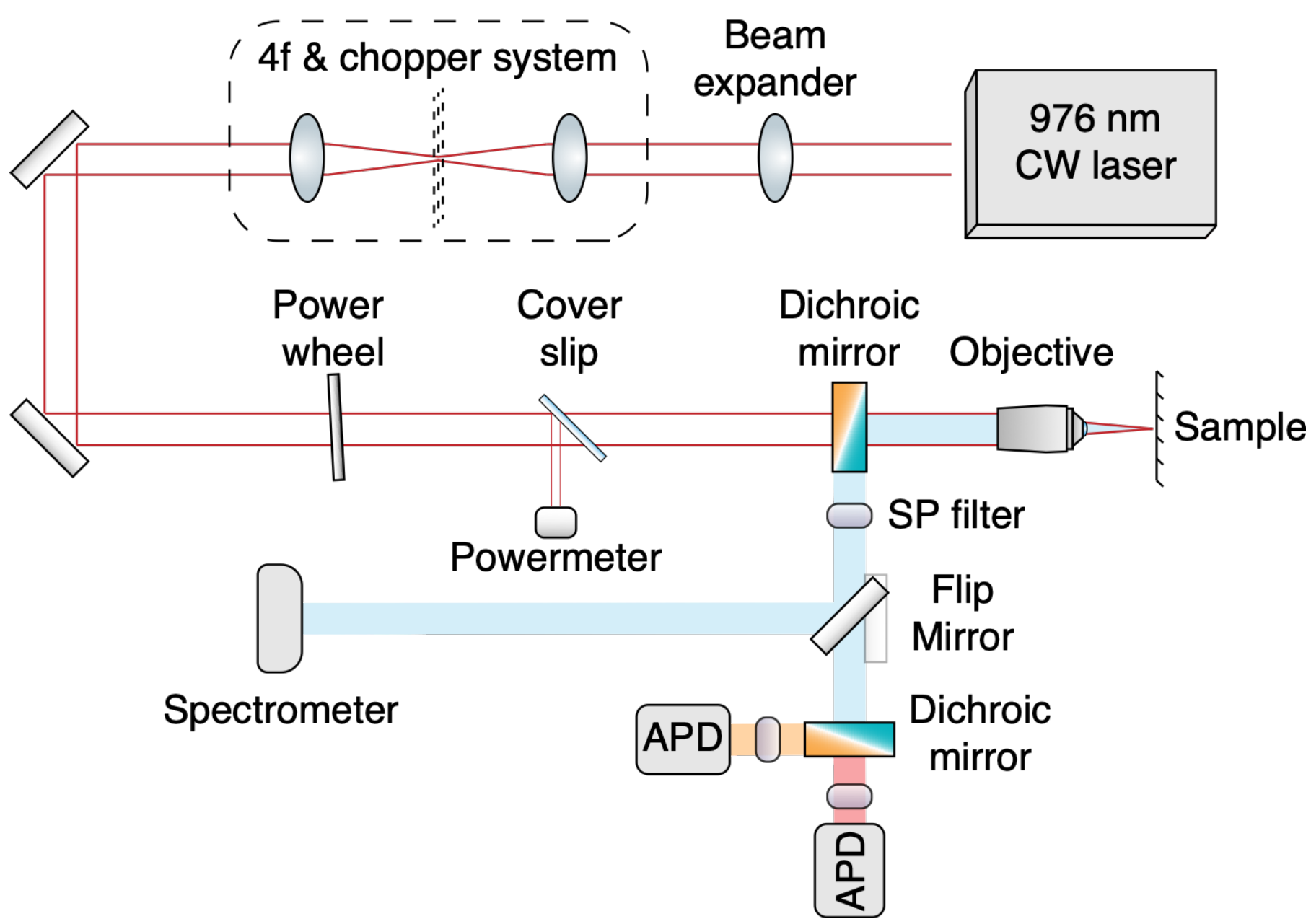

**Supplementary Scheme S1**. Scheme of the custom-built confocal laser scanning inverted microscope, equipped with a 976 nm CW laser, automated power adjustment, optical chopper, spectrometer, and a pair of avalanche photodiodes (APD).

### 1.9 Analysis of luminescence hysteresis

To analyze photoluminescence hysteresis of $Ho^{3+}$-activated (nano)materials, we have estimated the switch-on threshold, hysteresis width, and switch-on contrast from the power-dependent photoluminescence plots (schematically shown as *Intensity* vs *Power density* in **Supplementary Scheme S2**) as reported before.[5]

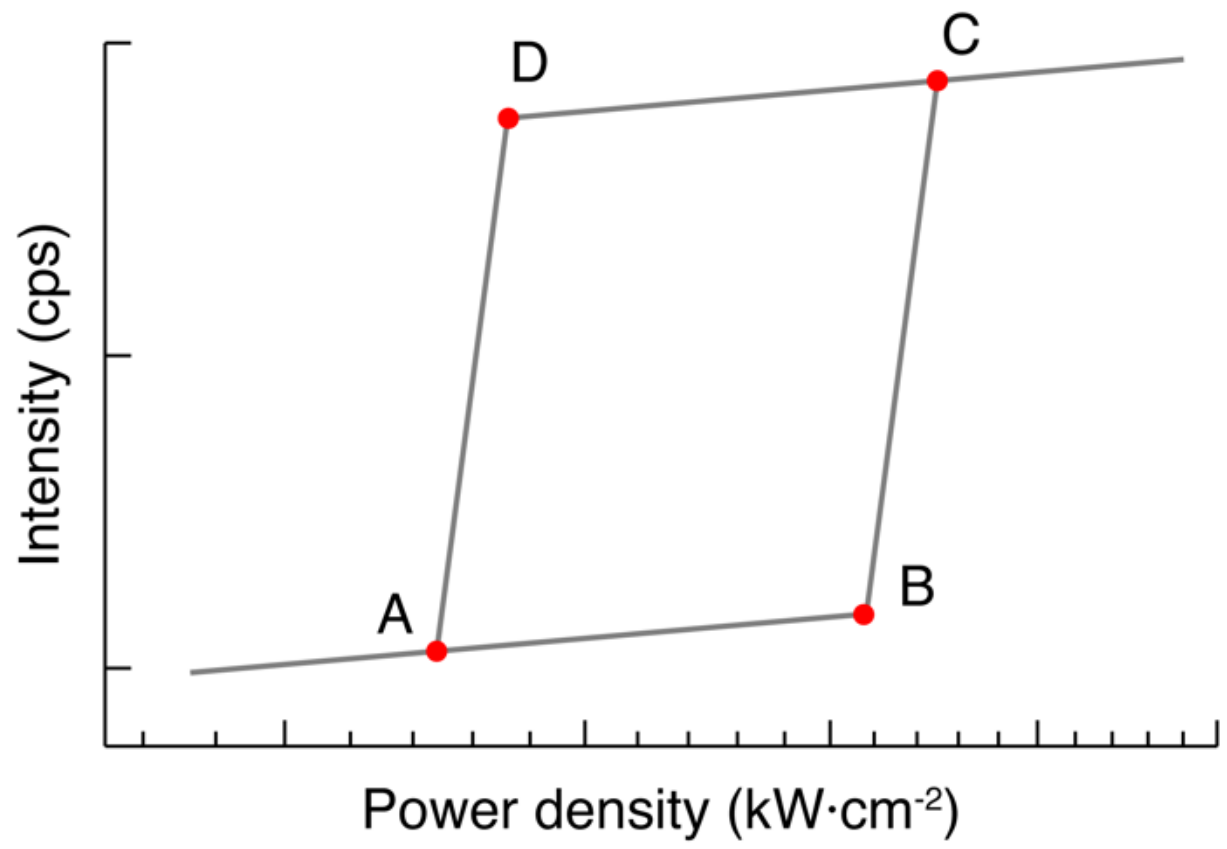

**Supplementary Scheme S2**. A schematic representation of an intensity (I) vs power density (P) graph, with key transition points of the hysteresis curve indicated by indices A, B, C, and D.

$$Switch-on\ threshold = P_B \tag{S1}$$

$$Hysteresis\ width = \frac{P_C + P_B}{2} - \frac{P_A + P_D}{2} \tag{S2}$$

$$Switch-on\ contrast = 10 * log_{10}\left[\frac{(I_C - I_B)}{I_{noise}}\right] \tag{S3}$$

Here, $I_{noise}$ = 160 cps, the average dark noise intensity attainable with our setup.

**Supplementary Section 2. Structural characterization of $Cs_2NaInCl_6:Ho^{3+}$ nanocrystals**

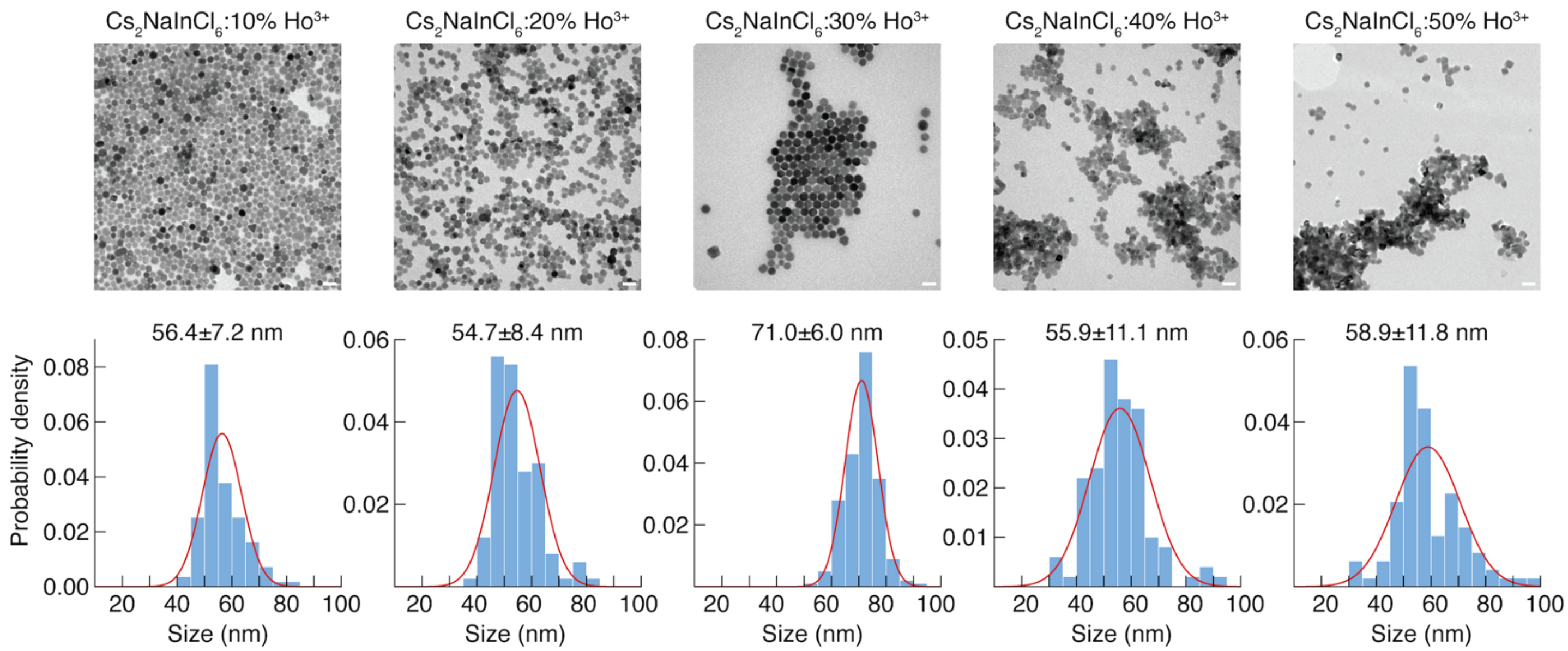

**Supplementary Figure S1**. TEM images and size distribution of $Cs_2NaInCl_6$: x% $Ho^{3+}$ (x = 10, 20, 30, 40, 50) nanocrystals. Scale bars are 100 nm.

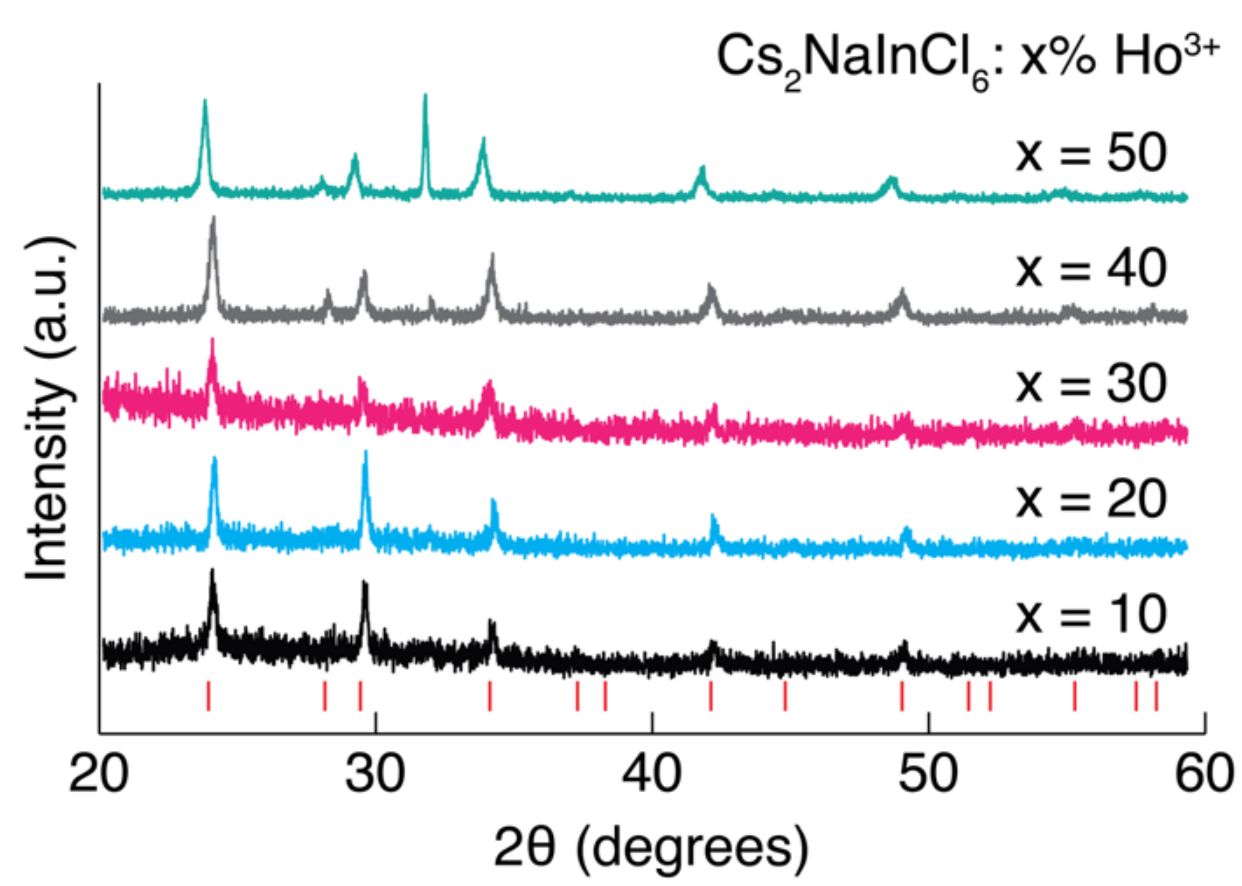

**Supplementary Figure S2**. XRD patterns of $Cs_2NaInCl_6$: x% $Ho^{3+}$ (x = 10, 20, 30, 40, 50) ANPs. The reference pattern for $Cs_2NaInCl_6$ was calculated based on the refined data provided in ref. 6 and is shown as red dashes.

**Supplementary Table S1**. Nominal vs actual concentration of $Ho^{3+}$ ions in $Cs_2NaInCl_6{:}Ho^{3+}$ nanocrystals with varied $Ho^{3+}$ mol% concentration. *concentration could not be detected reliably.

| $Cs_2NaInCl_6{:}Ho^{3+}$ sample: | Nominal $Ho^{3+}$ concentration, mol% | Actual $Ho^{3+}$ concentration, mol% |
|---|---|---|
| 1 | 10 | <5 * |
| 2 | 20 | <5 * |
| 3 | 30 | 18 |
| 4 | 40 | 31 |
| 5 | 50 | 42 |

**Supplementary Section 3. Optical characterization of $Cs_2NaInCl_6:Ho^{3+}$ nanocrystals**

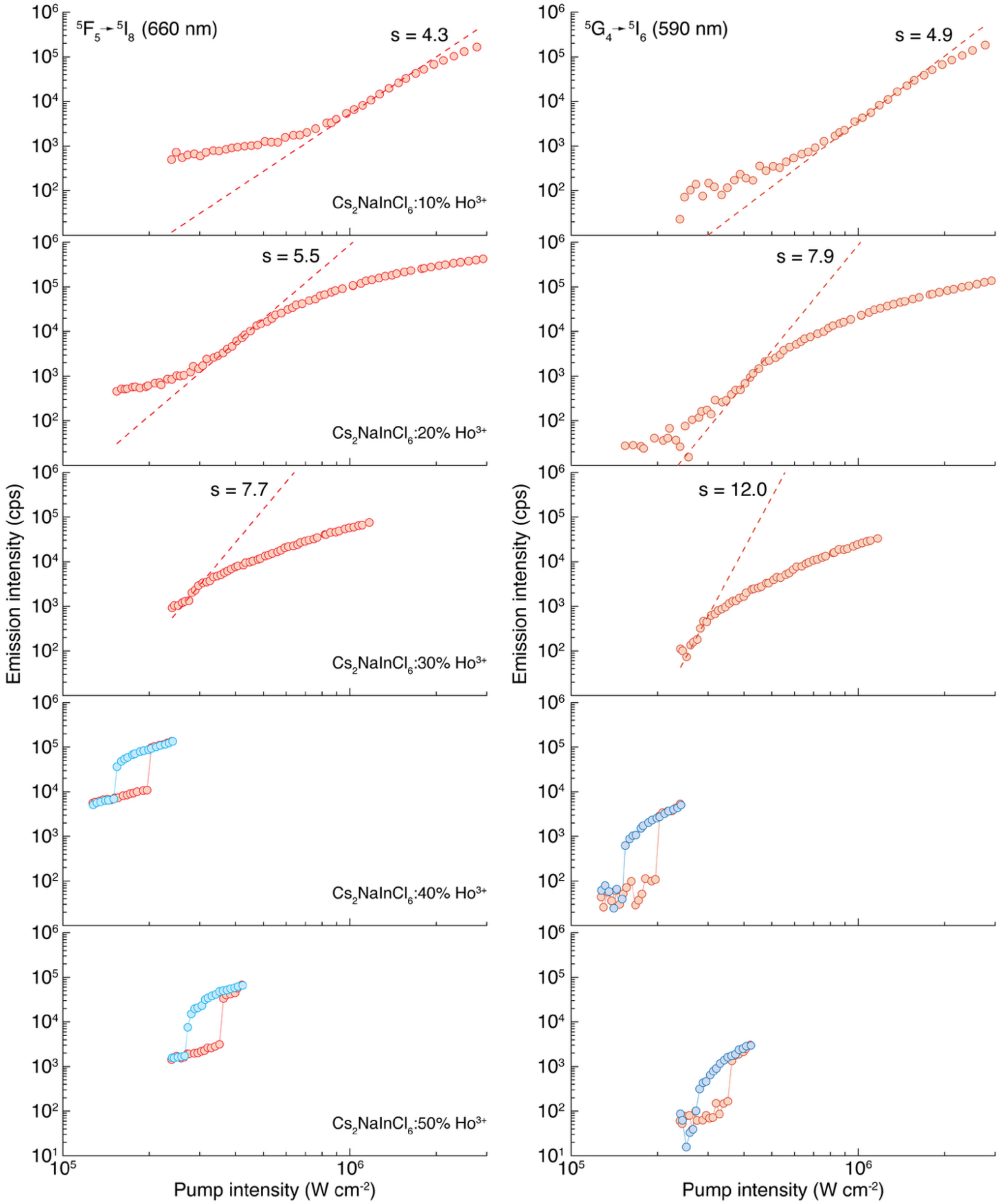

**Supplementary Figure S3**. 976 nm pump power dependence of $Ho^{3+}$ emission at 590 nm ($^5G_4 \rightarrow {}^5I_6$; right column) and 660 nm ($^5F_5 \rightarrow {}^5I_8$; left column) from $Cs_2NaInCl_6$: x% $Ho^{3+}$ ANPs at room temperature. For $Cs_2NaInCl_6$: x% $Ho^{3+}$ (x = 40 and 50) ANPs, red circles represent the PL intensity acquired with increasing excitation intensity (forward scan) and blue ones with decreasing excitation intensity (reverse scan). We detected no luminescence hysteresis in x = 10, 20, 30% $Ho^{3+}$-activated $Cs_2NaInCl_6$ nanocrystals; instead, linear fits to the log-log data indicate that the upconversion nonlinearity (*s*) increases with doping concentration.

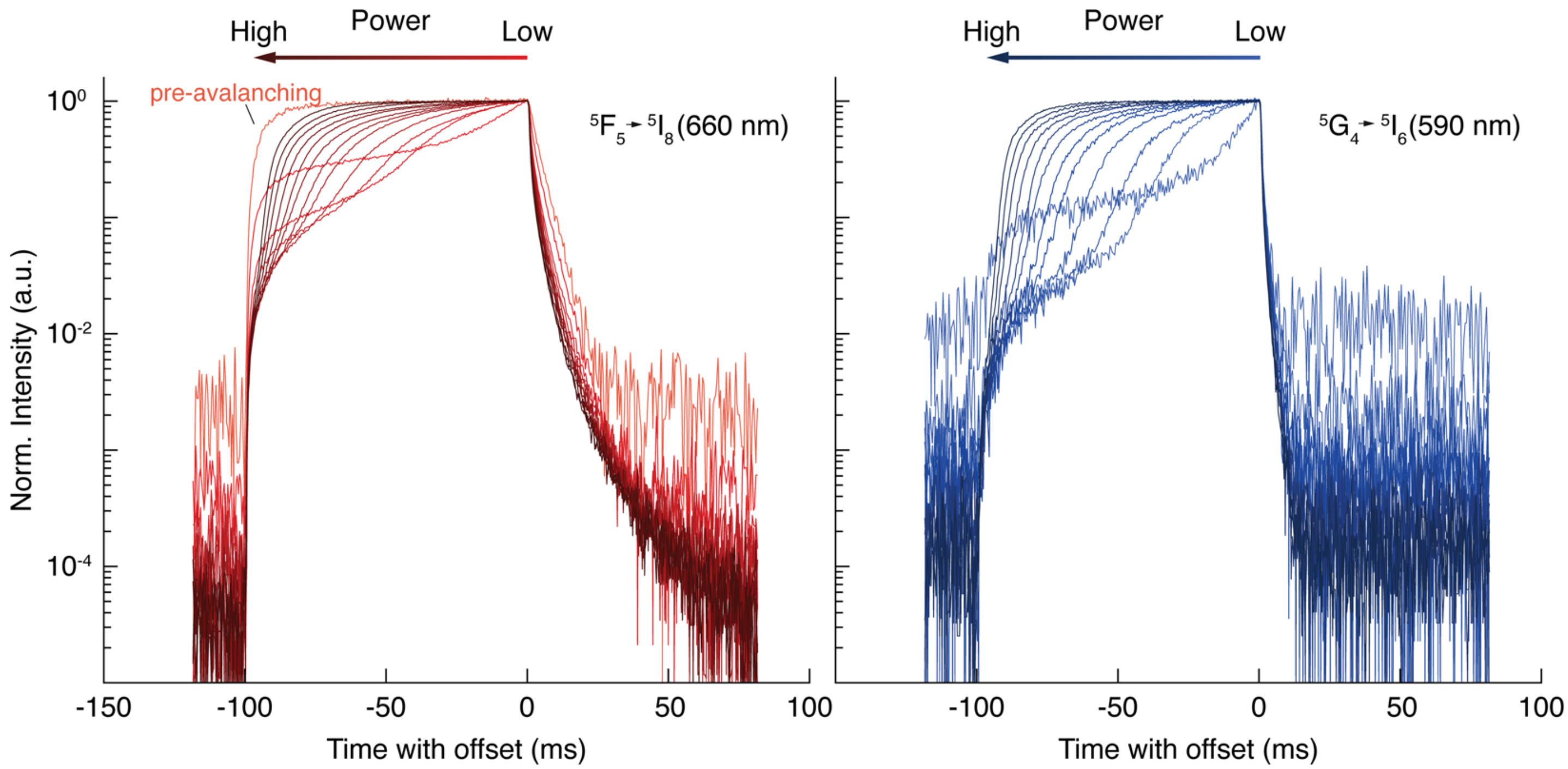

**Supplementary Figure S4**. Time-resolved photoluminescence of $Cs_2NaInCl_6$: 40% $Ho^{3+}$ ANPs at different 976 nm pump intensities. Right – emission acquired for the 590 nm channel ($^5G_4 \rightarrow {}^5I_6$); left – for the 660 nm channel ($^5F_5 \rightarrow {}^5I_8$). The figure also shows time-resolved luminescence of the 660 nm emission in the pre-avalanching regime. Traces are time-offset to distinguish rise and decay profiles relative to the falling edge of the excitation pulse. The 976 nm pump was pulsed at 4 Hz with a 50% duty cycle.

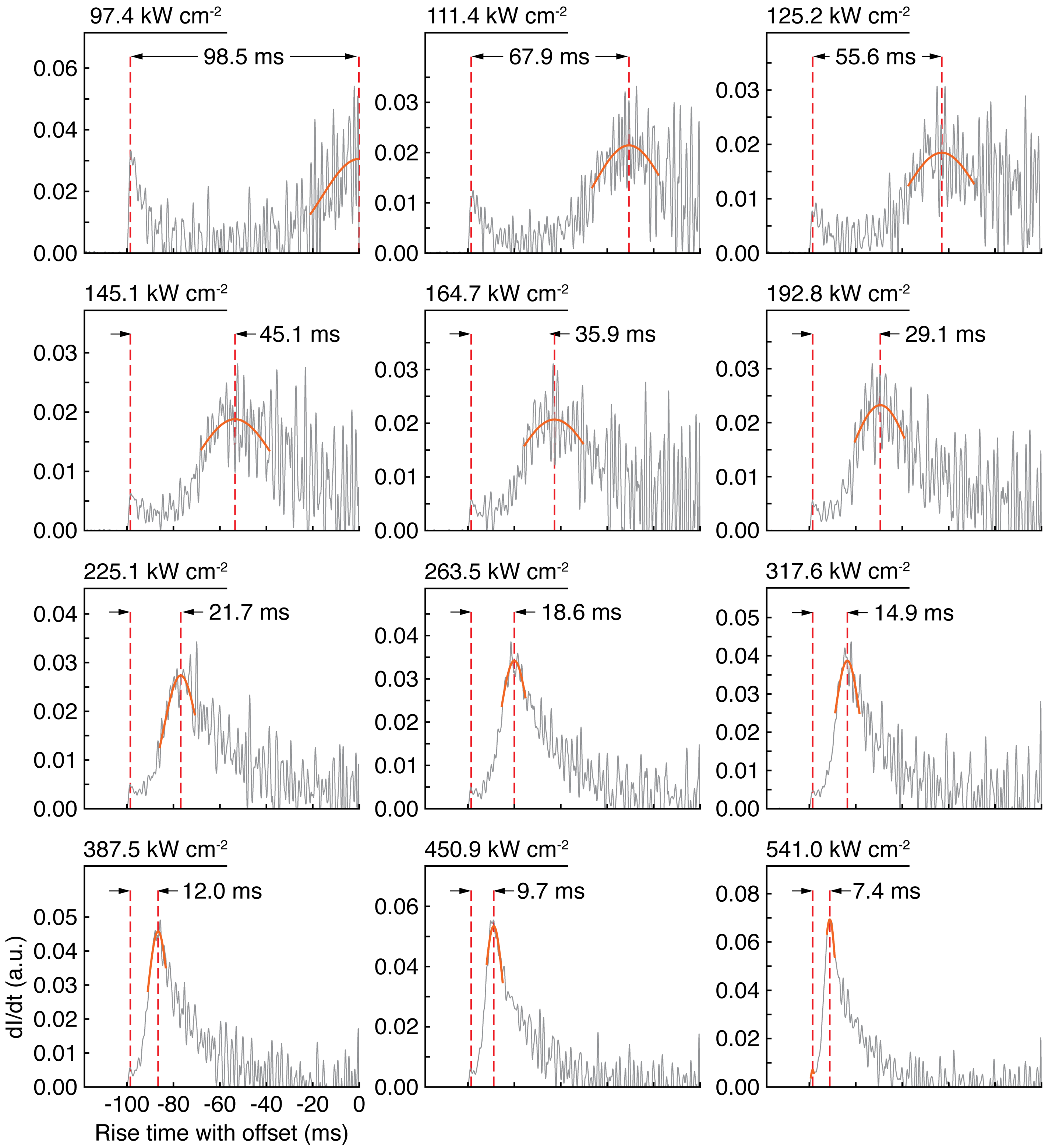

**Supplementary Figure S5**. Derivatives (*dI/dt*) of rise-profile data for 660 nm emission of $Cs_2NaInCl_6$: 40% $Ho^{3+}$ nanocrystals, as in **Supplementary Figure S4**. To improve the signal-to-noise ratio, the raw time-resolved data were first binned and smoothed using a Gaussian filter and a Savitzky-Golay filter (polynomial order 3). Peak values of *dI/dt* data were Gaussian-fitted to determine the extrema of the rise profiles. The corresponding average pump power intensities and the delay between the first and second rise-profile extrema are indicated. All graphs share the same x-axis values, offset relative to the falling edge of the excitation pulse.

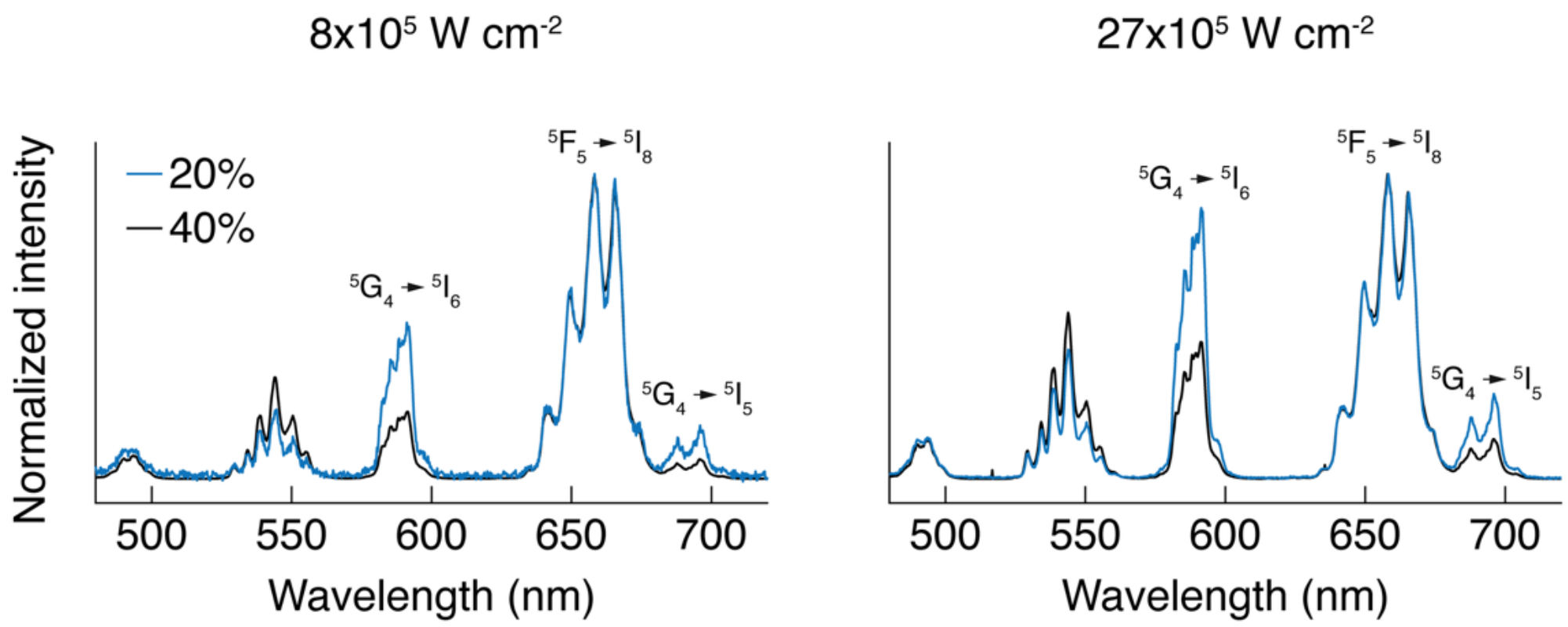

**Supplementary Figure S6**. Normalized upconversion emission of $Cs_2NaInCl_6$: x% $Ho^{3+}$ (x = 20 and 40) nanocrystals under different 976 nm pump intensities.

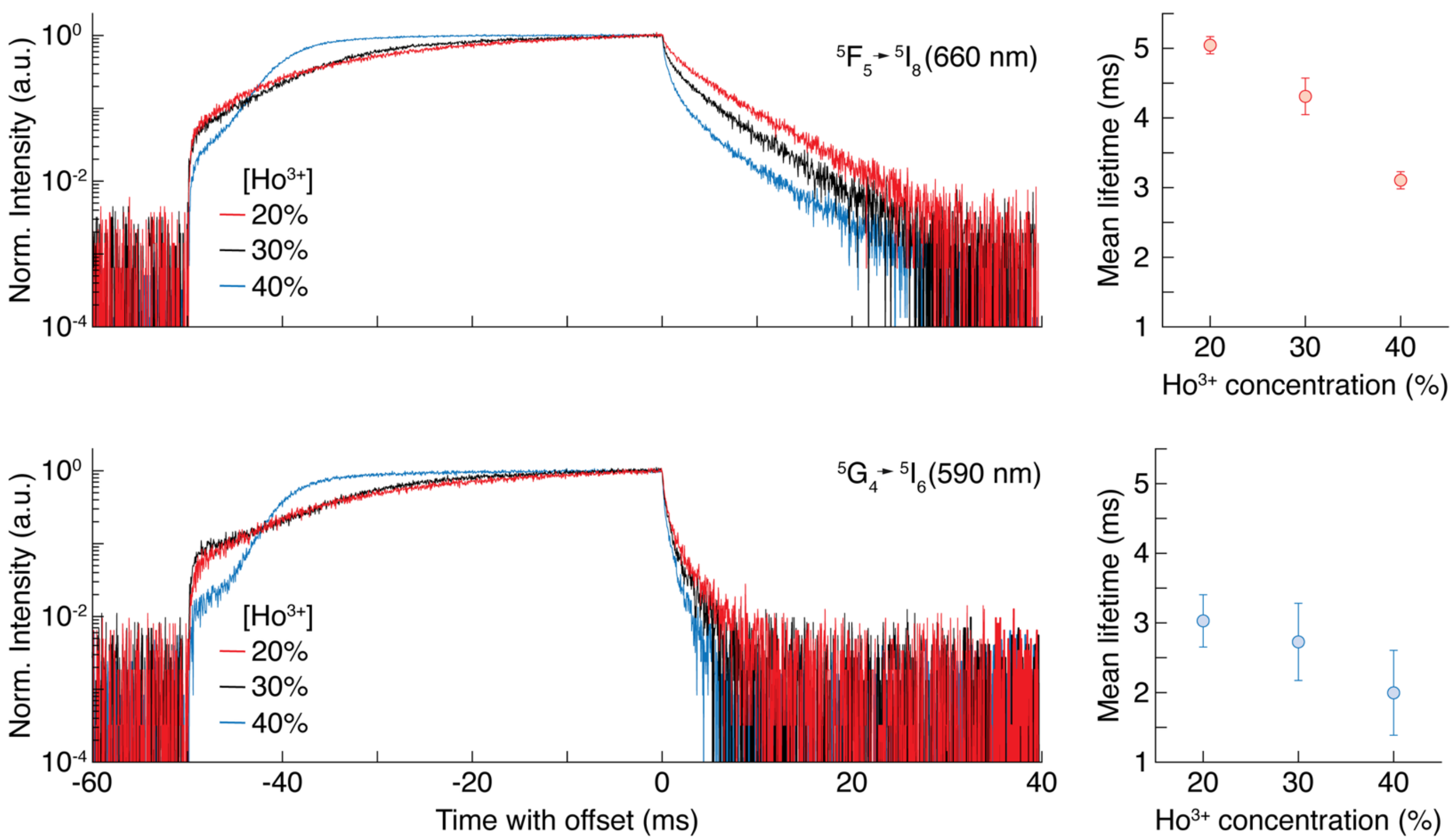

**Supplementary Figure S7**. Time-resolved photoluminescence of $Cs_2NaInCl_6$: x% $Ho^{3+}$ nanocrystals (x = 20, 30, and 40) at an average pump intensity of 600 kW·cm$^{-2}$. Bottom – emission acquired for the 590 nm channel ($^5G_4 \rightarrow {}^5I_6$); top – for the 660 nm channel ($^5F_5 \rightarrow {}^5I_8$). Traces are time-offset to distinguish rise and decay profiles relative to the falling edge of the excitation pulse. The 976 nm pump was pulsed at 4 Hz with a 50% duty cycle. $Ho^{3+}$ concentration-dependent lifetime values for the two emission channels are reported as mean + 1 std. dev. from 5 independent measurements across nanocrystal films. Mean lifetimes were calculated as the area under the curve of the decay profile.

**Supplementary Section 4. Structural and optical characterization of $NaYF_4:Ho^{3+}$ and $LiYF_4:Ho^{3+}$ nanoparticles**

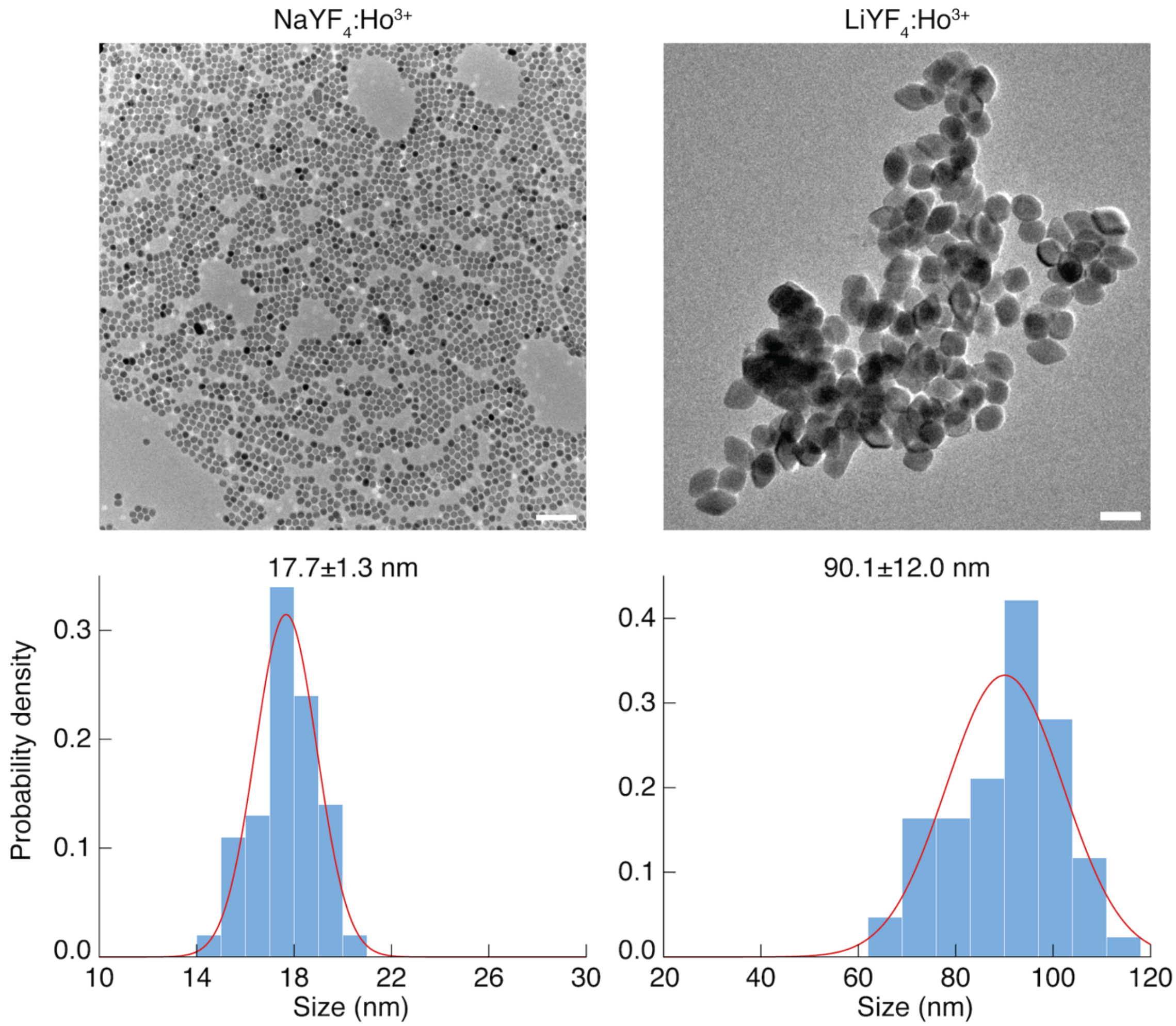

**Supplementary Figure S8**. TEM images and size distribution of $NaYF_4$: 20% $Ho^{3+}$ (left) and $LiYF_4$: 20% $Ho^{3+}$ (right) nanocrystals. Scale bars are 100 nm.

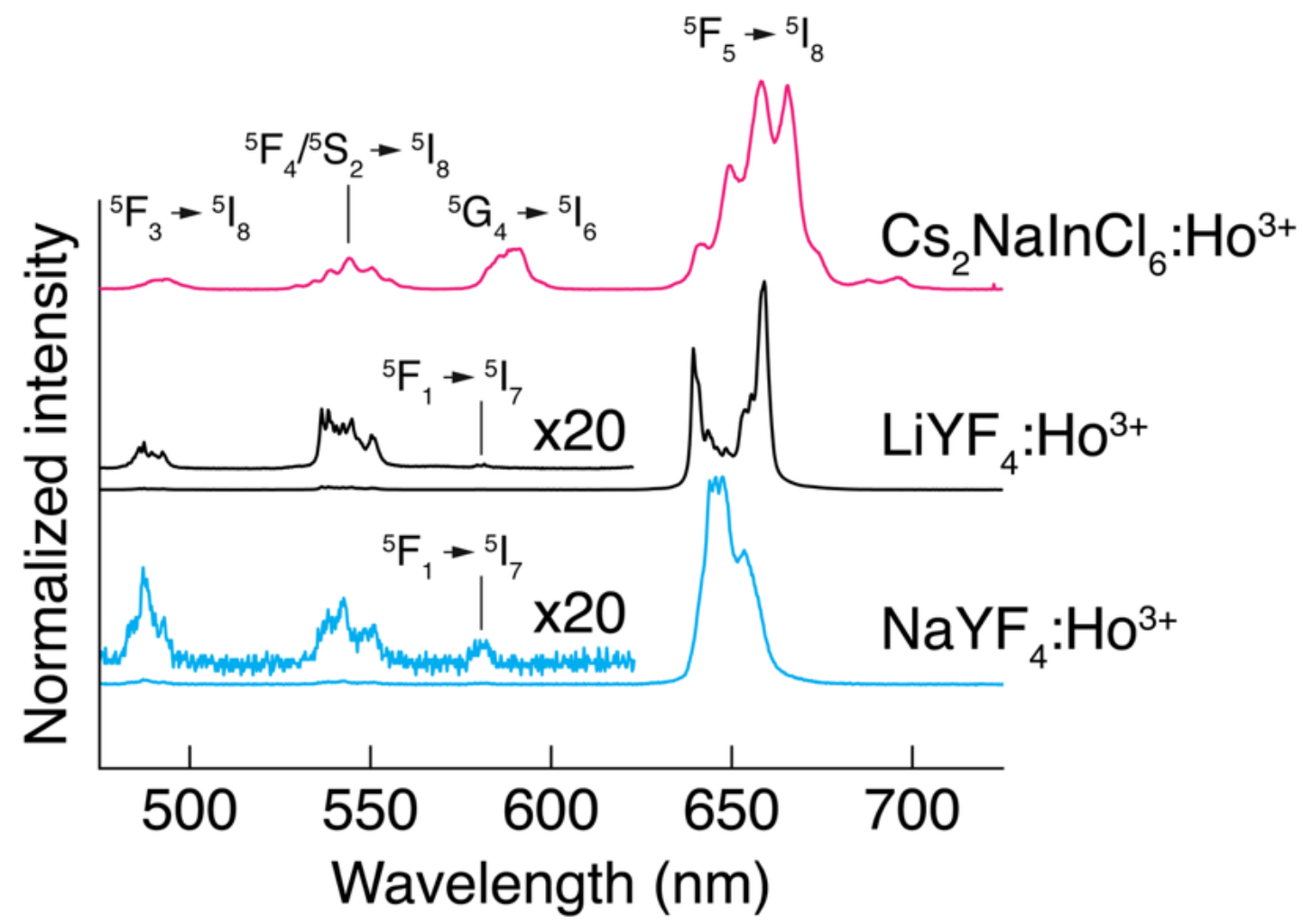

**Supplementary Figure S9**. Photoluminescence spectra of $Cs_2NaInCl_6:Ho^{3+}$, $NaYF_4:Ho^{3+}$, and $LiYF_4:Ho^{3+}$ nanocrystals under a 976 nm pump of 1 MW·cm$^{-2}$. Signal intensity of fluoride nanocrystals in the 475-625 nm range is multiplied by 20-fold for visibility.

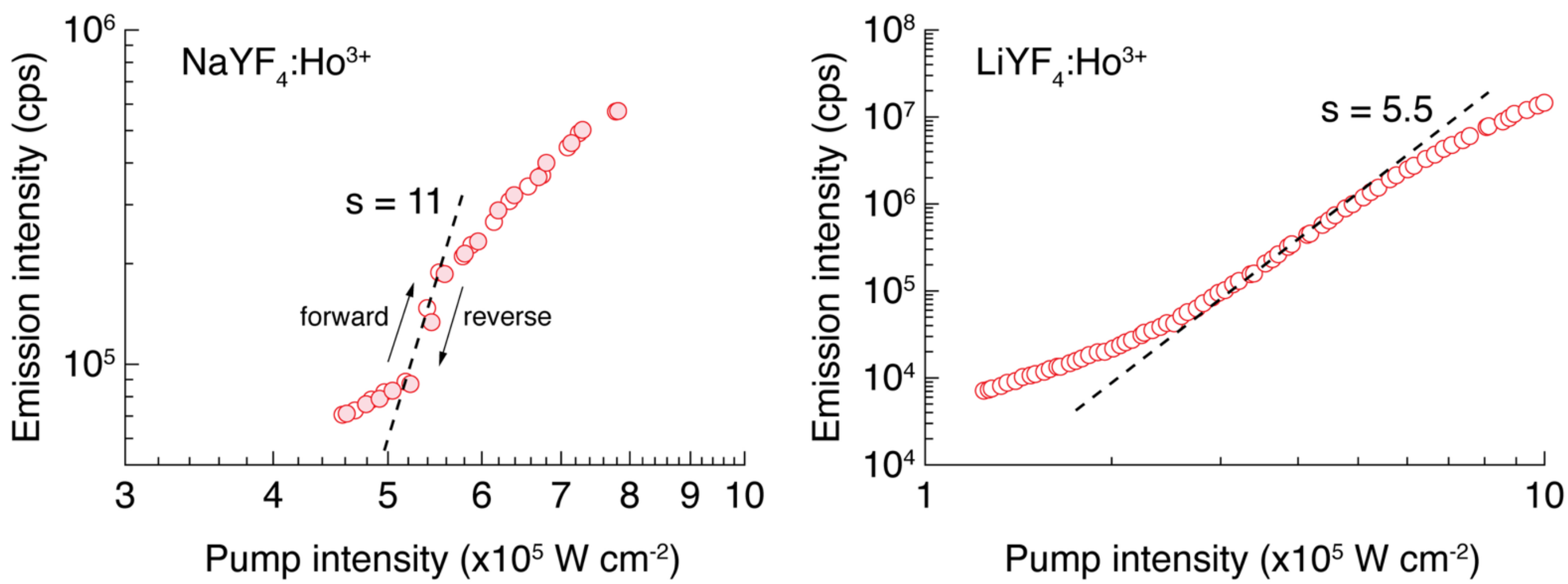

**Supplementary Figure S10**. 976 nm pump power dependence of $Ho^{3+}$ emission at 660 nm ($^5F_5 \rightarrow {}^5I_8$) from $NaYF_4:Ho^{3+}$ (left) and $LiYF_4:Ho^{3+}$ (right) nanocrystals at room temperature. For $NaYF_4:Ho^{3+}$ nanocrystals, closed circles show the PL intensity as excitation intensity increases (forward scan), and open circles show it as excitation intensity decreases (reverse scan). Linear fits to the log-log data indicate upconversion nonlinearity (*s*) of energy-looping/photon avalanching.

**Supplementary Section 5. Rate equation modeling of $Ho^{3+}$ photoactivation dynamics**

To the best of our knowledge, phenomenological Judd-Ofelt intensity parameters for the $Cs_2NaInCl_6:Ho^{3+}$ are not readily available. Therefore, we estimated these parameters following the JOEX method reported by A. Ćirić et al.[7] Using the dissolution-recrystallization synthesis, we prepared phase-pure $Cs_2NaInCl_6$: 5% $Eu^{3+}$ powders and measured their excitation spectrum at 615 nm (**Supplementary Figure S11b**).

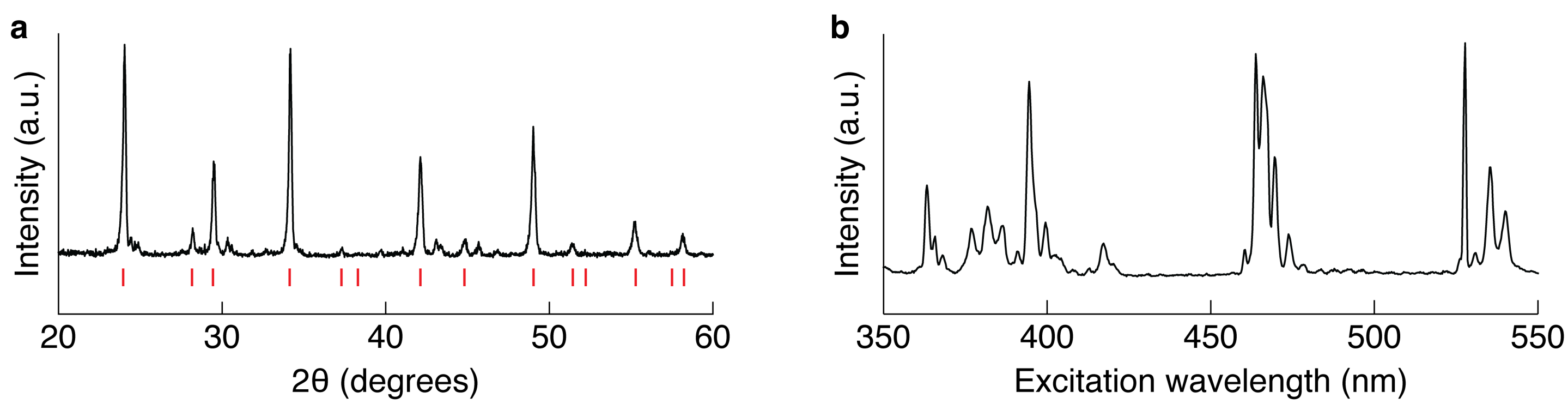

**Supplementary Figure S11. a**, XRD pattern of $Cs_2NaInCl_6:Eu^{3+}$ microcrystals. The reference pattern for $Cs_2NaInCl_6$ was calculated based on the refined data provided in ref. 6 and is shown as red dashes. **b**, Photoluminescence excitation spectrum of $Cs_2NaInCl_6:Eu^{3+}$, recorded by monitoring emission at 615 nm ($^5D_0 \rightarrow {}^7F_2$).

**Supplementary Table S2.** Judd-Ofelt intensity parameters ($\Omega_\lambda$) estimated from the excitation spectrum of $Cs_2NaInCl_6:Eu^{3+}$. *refractive indices were estimated from I. Khan et al.[8]

| λ | n* | Center wavelength | Integrated Intensity | $\Omega_\lambda \cdot 10^{-20}$ (cm$^2$) |
|---|---|---|---|---|
| 2 | 1.6958 | 364.1 | 0.82 | **3.038** |
| 4 | 1.6810 | 395.3 | 1.86 | **0.448** |
| 6 | 1.6605 | 464.9 | 3.46 | **0.080** |
| MD | 1.6520 | 527.6 | 0.95 | |

Note that intensity parameters ($\Omega_\lambda$) estimated in **Supplementary Table S2** are effective parameters that encode a mixture of true Judd-Ofelt parameters and vibronic coupling, due to the centrosymmetric nature of the double perovskite host. Therefore, we use these parameters qualitatively and supplement the $Ho^{3+}$ photoactivation dynamics model with the better-established $KPb_2Cl_5$ heavy-halide host, which serves as a general surrogate for a low-phonon-energy host model.

**Theoretical description of the system and photoactivation dynamics modeling**

Since the intensity of a given radiative transition is proportional to the product of the emitting-state population and the microscopic transition rate constant, we used a set of coupled differential equations to describe the instantaneous populations of the $Ln^{3+}$ manifolds[9]. In this population balance model, the population $N_i$ (in $nm^{-3}$) of a lanthanide $4f^N$ manifold $i$ over time is determined by the incoming and outgoing rates of electric dipole (ED) and magnetic dipole (MD) radiative transitions, nonradiative multiphonon relaxation (MPR), and energy transfer (ET):

$$\frac{dN_i}{dt} = \sum_j \left(N_j A_{ji}^{ED} - N_i A_{ij}^{ED}\right) + \sum_j \left(N_j A_{ji}^{MD} - N_i A_{ij}^{MD}\right) + \left(N_{i+1} W_{i+1,i}^{NR} - N_i W_{i,i-1}^{NR}\right) + \sum_{ij,kl} \left(N_j N_l P_{ji,kl}^{ET} - N_i N_k P_{ij,lk}^{ET}\right) \quad \text{(S4)}$$

$A_{ij}^{ED}$ and $A_{ij}^{MD}$ ($s^{-1}$) are the Einstein coefficients for ED and MD radiative transitions from manifold $i$ to $j$. $W_{i,i-1}^{NR}$ ($s^{-1}$) is the nonradiative MPR rate from the manifold $i$ to the manifold immediately below $i$. $P_{ij,kl}^{ET}$ ($nm^3 \cdot s^{-1}$) is the microscopic energy transfer parameter for the transfer of energy for the donor $i$ to $j$ transition and the acceptor $k$ to $l$ transition.

Transition probabilities for individual lanthanide ions are calculated using the same theoretical framework as the comprehensive rate equation models described in our previous work[9,10] (see ref. [11] for the most detailed description). The transitions in trivalent lanthanide ions are $4f^N$ manifold-to-manifold transitions, including single-ion transitions (non-radiative multi-phonon relaxation, electric dipole absorption/emission, and magnetic dipole absorption/emission) and two-ion interactions (dipole-dipole energy transfer). The individual theories and equations behind these transition probabilities are discussed below.

MPR transition probabilities, $W_i^{MPR}$, from an initial lanthanide manifold $i$ to the level below $i$ are calculated using van Dijk's modified Energy Gap Law[12]:

$$W_i^{NR} = W_0^{MPR} e^{-\alpha(\Delta E - 2\hbar\omega)} \quad \text{(S5)}$$

where $\Delta E$ is the energy gap to the lower manifold and $\hbar\omega$ is the energy of the cutoff phonon energy of the host matrix. $W_0^{MPR}$ and $\alpha$ are MPR constants that are consistent across transitions in a given host matrix. $W_0^{MPR}$ is the MPR rate when $\Delta E = 2\hbar\omega_{max}$.

Electric dipole (ED) radiation and absorption probabilities are calculated using Judd-Ofelt (JO) theory[11,13]. The transition probability (in $s^{-1}$) for electric dipole radiative relaxation, $A_{ij}^{ED}$, is determined using the three JO parameters ($\Omega_\lambda$) and the doubly reduced matrix elements of the unit tensor operator $U^{(\lambda)}$ of rank $\lambda$, where $\lambda =$ 2,4, or 6:

$$A_{ij}^{ED} = \frac{64\pi^4 e^2 \tilde{\upsilon}^3}{3h(2J+1)}\left[\frac{n(n^2+2)^2}{9}\right] S_{ij}^{ED} \quad \text{(S6)}$$

$$S_{ij}^{ED} = \sum_{\lambda=2,4,6} \Omega_\lambda |< [SL]J_i || U^{(\lambda)} || [SL]J_j >|^2 \quad \text{(S7)}$$

Here, $\tilde{\upsilon}$ is the energy of the transition in wavenumbers, $h$ is Planck's constant, $e$ is the elementary charge, $n$ is the index of refraction of the material, $J$ is the total angular momentum of the initial state, and $S_{ij}^{ED}$ is the electric dipole line strength, using the Gaussian unit system. JO parameters used in this work are shown in **Supplementary Table S3**.

**Supplementary Table S3.** Judd-Ofelt parameters and reduced matrix elements. *Judd-Ofelt intensity parameters estimated from the photoluminescence excitation spectrum of $Cs_2NaInCl_6:Eu^{3+}$ are used as a proxy for simulating photoactivation dynamics of $Cs_2NaInCl_6:Ho^{3+}$.

| Parameter | $Eu^{3+}$ in $Cs_2NaInCl_6$* | $Ho^{3+}$ in $KPb_2Cl_5$ | $Ho^{3+}$ in fluoride |
|---|---|---|---|
| $\boldsymbol{\Omega}_2$ ($10^{-20}$ $cm^2$) | 3.038 | 1.30 | 1.16 |
| $\boldsymbol{\Omega}_4$ | 0.448 | 1.34 | 1.62 |
| $\boldsymbol{\Omega}_6$ | 0.080 | 1.06 | 1.60 |
| Source, $\boldsymbol{\Omega}_\lambda$ | Ćirić et al.[7] | Sardar et al.[14] | Kaminskii et al.[15] |
| Source, $\|\langle i\|U_\lambda\|j\rangle\|^2$ | Kaminskii et al.[15] | | |

For ED absorption, $A_{ij} = \psi\sigma_{ij}$, where $\psi$ is the incident photon flux. The ED absorption cross section integrated over the transition, $\sigma_{ij}^{ED}$, is calculated from $S_{ij}^{ED}$ using the relation:

$$\sigma_{int}^{ED} = \frac{8\pi^3 e^2 \tilde{\upsilon}}{3hcg_i}\frac{(n^2+2)^2}{9n}S_{ij}^{ED} \quad \text{(S8)}$$

Here, $\sigma_{int}^{ED}$ is the ED absorption cross section integrated over the entire transition. We handled two cases: resonant and non-resonant (phonon-assisted) absorption. For resonant absorption, we assumed that the laser excitation source has a width of 1 nm, and the absorption peak is a Gaussian curve centered at $\tilde{\upsilon}$. In this case, $\sigma_{ij}^{ED}$ is determined from $\sigma_{int}^{ED}$ using the expression:

$$\sigma_{ij} = \sigma_{int}^{ED}\frac{1}{w\sqrt{2\pi}}e^{-(\upsilon_{ex}-\tilde{\upsilon})^2/(2w^2)} \quad \text{(S9)}$$

where $\upsilon_{ex}$ is the energy (in wavenumbers) of the excitation source. $w$ is the linewidth of the absorption transition, related to the absorption FWHM by the expression: $w = FWHM_{abs}/(2\sqrt{2ln2})$. For $FWHM_{abs}$ see **Supplementary Table S4**.

For non-resonant absorption, the cross-section was modified to account for phonon assistance, as for MPR:

$$\sigma_{ij,PA} = \sigma_{ij,0}.e^{-\alpha|\upsilon_{ex}-\tilde{\upsilon}|} \quad \text{(S10)}$$

where $\sigma_{ij,0}$ is the value of the resonant absorption cross section when $\upsilon_{ex} = \tilde{\upsilon}$.

**Magnetic dipole (MD) radiation and absorption probabilities**. The MD radiative rate constant $A_{ij}^{MD}$ is calculated with the equations:

$$A_{ij}^{MD} = \frac{64\pi^4 e^2 \tilde{\upsilon}^3}{3h(2J+1)}n^3 S_{ij}^{MD} \quad \text{(S11)}$$

$$S_{ij}^{MD} = \mu_B^2 | < [SL]J_i || L + 2S || [SL]J_j > |^2 \quad \text{(S12)}$$

where $\mu_B$ is the Bohr magneton. $L$ and $S$ are the angular momentum and spin operators, respectively. Note: these operators must be used on the eigenvectors of the intermediate coupled states of each lanthanide ion since each manifold total angular momentum $J$ is treated as a linear combination of various $LS$ states with the same $J$. The integrated MD absorption cross-section and oscillator strength are calculated using the expressions below:

$$\sigma_{\text{int}}^{MD} = \frac{\pi e^2}{mc^2} f_{ij}^{MD} \quad \text{(S13)}$$

$$f_{ij}^{MD} = \frac{8\pi^2 mc\tilde{v}}{3he^2 g_i} n S_{ij}^{MD} \quad \text{(S14)}$$

Here, $\sigma_{int}^{MD}$ is the MD absorption cross-section integrated over the entire transition. The actual MD absorption probability is calculated from $\sigma_{int}^{MD}$ in the same manner described for ED absorption above.

**Phonon-assisted energy transfer probabilities.** The orientation-averaged rate of dipole-dipole energy transfer $W_{ij,kl}^{ET}$ with donor transition $i \rightarrow j$ and acceptor transition $k \rightarrow l$ is described as a function of the inter-ion distance $R$ and the ED line strengths $S^{ED}$ of the donor and acceptor transitions[16]:

$$W_{ij,kl}^{ET} = \frac{C_{DA,ijkl}}{R^6} \quad \text{(S15)}$$

$$C_{DA,ijkl} = \frac{8\pi^2 e^4 s_0}{3h^2 c g_i g_k} \left(\frac{n^2+2}{3n}\right)^4 S_{ij}^{ED} S_{kl}^{ED} \quad \text{(S16)}$$

In this expression, $C_{DA,ijkl}$ is the energy transfer micro-parameter, $g_i$ is the $2J+1$ number of states in manifold $i$, and $s_0$ is the overlap integral between the normalized donor emission spectrum and acceptor absorption spectrum. To account for phonon assistance during non-resonant energy transfer, the phonon-assisted energy transfer micro-parameter is used in place of the resonant ET micro-parameter and is calculated as[17]:

$$C_{DA}^{PAET} = C_{DA,ijkl} \cdot exp[-\beta_{MPR}\Delta E] \quad \text{(S17)}$$

where $\beta_{MPR} = \alpha - \ln(2)/\hbar\omega$, and $\Delta E$ is the net energy difference between the donor and acceptor transitions. $C_{DA,ijkl}$ is the energy transfer micro-parameter assuming a resonant ET process, calculated using Equation S16.

To account for energy transfer over all possible donor-acceptor distances *R* in a nanocrystal, the total phonon-assisted energy transfer (PAET) rate constant for a given donor can be calculated by integrating Equation S17 over all *R* (ref: [18]). In the case of a homogeneously doped material (see Supporting Information of ref: 11), we can define the PAET microparameter as:

$$W_{ij,kl,avg}^{PAET} = \frac{4\pi}{3} N_A C_{DA}^{PAET} R_{\min}^{-3} = N_A P_{ij,kl}^{MAET} \quad \text{(S18)}$$

To make the rate constant independent from the acceptor population, Equation S4 uses the constant, $P_{ij,kl}^{MAET}$ (units of $cm^3$), whose relationship with $W_{ij,kl,avg}^{MAET}$ is shown above.

We calculated the steady-state population of each $Ho^{3+}$ manifold under a given irradiation through numerical integration of the rate equations in Equation S4, which account for tens of thousands of simultaneous transitions. When performing power-dependent simulations, the population of $Ho^{3+}$ manifolds under a given power density value was simulated starting from the populations obtained at a prior power value.

Kinetic calculations were performed according to the method of Chan et al.[9,11] using Igor Pro 9 (Wavemetrics). N ordinary differential equations, which represent the population of each N manifold in the simulated system, were solved numerically using the Igor Pro Backward Differentiation Formula integration method. Calculation parameters are given in **Supplementary Table S4**.

**Supplementary Table S4.** General simulation parameters. *In $Cs_2NaInCl_6$:$Ho^{3+}$, volume per potential dopant site and minimum dopant distance were calculated assuming a possible $Na^+$ substitution for $Ho^{3+}$, as recently reported.[19] This correction requires experimental validation beyond the scope of this work, but it provides a geometric host description our model can simulate. Simulations without this assumption, or at very high $Ho^{3+}$ doping concentrations, failed to converge.

| Parameter | Value $Cs_2NaInCl_6$:$Ho^{3+}$ | Value $KPb_2Cl_5$:$Ho^{3+}$ | Value $NaYF_4$:$Ho^{3+}$ | Units |
|---|---|---|---|---|
| Simulation time period | 2500 | 2000 | 1000 | ms |
| Radiative rate ($A_{ij}$) threshold | $10^{-4}$ | $10^{-4}$ | $10^{-4}$ | $s^{-1}$ |
| Phonon energy[20,21] | 250 | 250 | 450 | $cm^{-1}$ |
| $W^0_{MPR}$ rate constant[12] | $0.5\cdot10^7$ | $0.5\cdot10^7$ | $1.0\cdot10^7$ | $s^{-1}$ |
| $\alpha$ MPR rate constant[12] | $15\cdot10^{-3}$ | $12\cdot10^{-3}$ | $3.5\cdot10^{-3}$ | cm |
| Index of refraction | 1.67 | 1.96 | 1.50 | - |
| Volume per potential dopant site* | 0.1555 | 0.1082 | 0.072395 | $nm^3$ |
| Minimum dopant distance* | 0.536 | 0.4765 | 0.3867 | nm |
| Absorption FWHM | 500 | 500 | 400 | $cm^{-1}$ |
| Incident laser power density | $10^5$-$10^7$ | $10^5$-$10^7$ | $10^5$-$10^7$ | $W\cdot cm^{-2}$ |
| Incident excitation wavelength | 976 | 976 | 976 | nm |

Cross-relaxation knock-out experiments were performed by ignoring transitions that met specified *i→j & l→k* criteria (**Supplementary Table S5**) during the assembly of rate equations (Equation S4)[10].

**Supplementary Table S5.** Summary of single and double cross-relaxation pathway knockouts (CR1-CR4) used in the rate-equation model. The following convention is used: [*i*, *l*] → [*j*, *k*]

| Knockout | Type | Transitions removed | Critical pathways remaining |
|---|---|---|---|
| CR1 | Single | CR1: [$^5F_3$, $^5I_8$] → [$^5F_5$, $^5I_7$] | CR2, CR3, CR4 |
| CR2 | Single | CR2: [$^5F_3$, $^5I_8$] → [$^5I_7$, $^5F_5$] | CR1, CR3, CR4 |
| CR3 | Single | CR3: [$^5G_4$, $^5I_8$] → [$^5F_3$, $^5I_7$] | CR1, CR2, CR4 |
| CR4 | Single | CR4: [$^5F_4$/$^5S_2$, $^5I_8$] → [$^5I_4$, $^5I_7$] | CR1, CR2, CR3 |
| CR1+CR2 | Double | CR1: [$^5F_3$, $^5I_8$] → [$^5F_5$, $^5I_7$]<br>CR2: [$^5F_3$, $^5I_8$] → [$^5I_7$, $^5F_5$] | CR3, CR4 |
| CR1+CR3 | Double | CR1: [$^5F_3$, $^5I_8$] → [$^5F_5$, $^5I_7$]<br>CR3: [$^5G_4$, $^5I_8$] → [$^5F_3$, $^5I_7$] | CR2, CR4 |
| CR1+CR4 | Double | CR1: [$^5F_3$, $^5I_8$] → [$^5F_5$, $^5I_7$]<br>CR4: [$^5F_4$/$^5S_2$, $^5I_8$] → [$^5I_4$, $^5I_7$] | CR2, CR3 |
| CR2+CR3 | Double | CR2: [$^5F_3$, $^5I_8$] → [$^5I_7$, $^5F_5$]<br>CR3: [$^5G_4$, $^5I_8$] → [$^5F_3$, $^5I_7$] | CR1, CR4 |
| CR2+CR4 | Double | CR2: [$^5F_3$, $^5I_8$] → [$^5I_7$, $^5F_5$]<br>CR4: [$^5F_4$/$^5S_2$, $^5I_8$] → [$^5I_4$, $^5I_7$] | CR1, CR3 |
| CR3+CR4 | Double | CR3: [$^5G_4$, $^5I_8$] → [$^5F_3$, $^5I_7$]<br>CR4: [$^5F_4$/$^5S_2$, $^5I_8$] → [$^5I_4$, $^5I_7$] | CR1, CR2 |

To determine critical transitions for a given energy transfer pathway, we considered only transitions with the highest branching fractions ($\beta_{ij}$) or contribution fractions ($\kappa_{ij}$), where

$$\beta_{ij,t} = \frac{\left(\frac{dN_i}{dt}\right)_{i\to j}^{t}}{\sum_{t=ED,MD,ET,MPR} \sum_m \left(\frac{dN_i}{dt}\right)_{i\to m}^{t}} \quad \text{(S19)}$$

$$\kappa_{ij,t} = \frac{\left(\frac{dN_j}{dt}\right)_{i\to j}^{t}}{\sum_t \sum_m \left(\frac{dN_j}{dt}\right)_{m\to j}^{t}} \quad \text{(S20)}$$

For a given transition type $t$ (ED/MD radiative, ET, MPR) from $i$ to $j$, $\kappa_{ij}$ ($\beta_{ij}$) represents that transition's fractional contribution to the overall population (depletion) of $j$ ($i$) by all types of transitions to (from) level $j$ ($i$). The rates $dN_i/dt$ are determined by the populations of the originating states and the corresponding microscopic rate constants, as shown by the individual components of Equation S4. To select the most significant incoming transitions for each manifold $i$, we retained up to 10 transitions with the highest $\kappa_{ij}$ that contributed to the respective population of level $j$. The most significant depopulating transitions were filtered similarly using $\beta_{ij}$ values.

The full set of pathway analysis results is provided in **Supplementary Data 1**. It includes contributing and branching ratios, rate constants, and rates for both the $Cs_2NaInCl_6{:}Ho^{3+}$ and $KPb_2Cl_5{:}Ho^{3+}$ systems evaluated at two bistable points corresponding to the forward and reverse power sweeps.

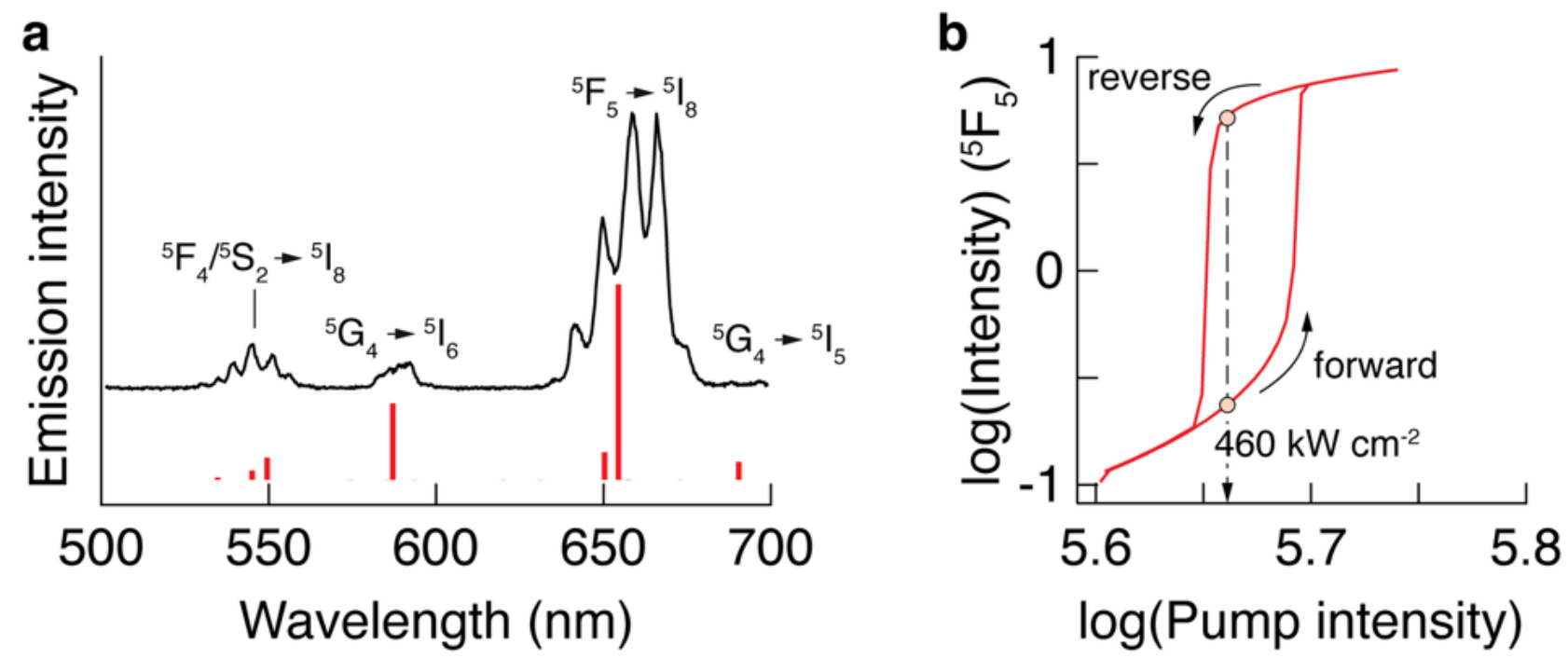

**Supplementary Figure S12**. **a**, Experimental (black) and simulated (red) emission spectra of $Cs_2NaInCl_6$:$Ho^{3+}$ materials. **b**, Simulated pump (976 nm) power dependence of $Ho^{3+}$ emission (660 nm, $^5F_5 \rightarrow {}^5I_8$) in $Cs_2NaInCl_6$: 20% $Ho^{3+}$. Both forward (increasing power) and reverse (decreasing power) scans are present, indicating luminescence hysteresis across the entire power scan. Circles around 460 kW·cm$^{-2}$ represent two steady states for which additional pathway analysis (**Supplementary Data 1**) and mechanism elaboration (**Supplementary Figure S13**) were performed.

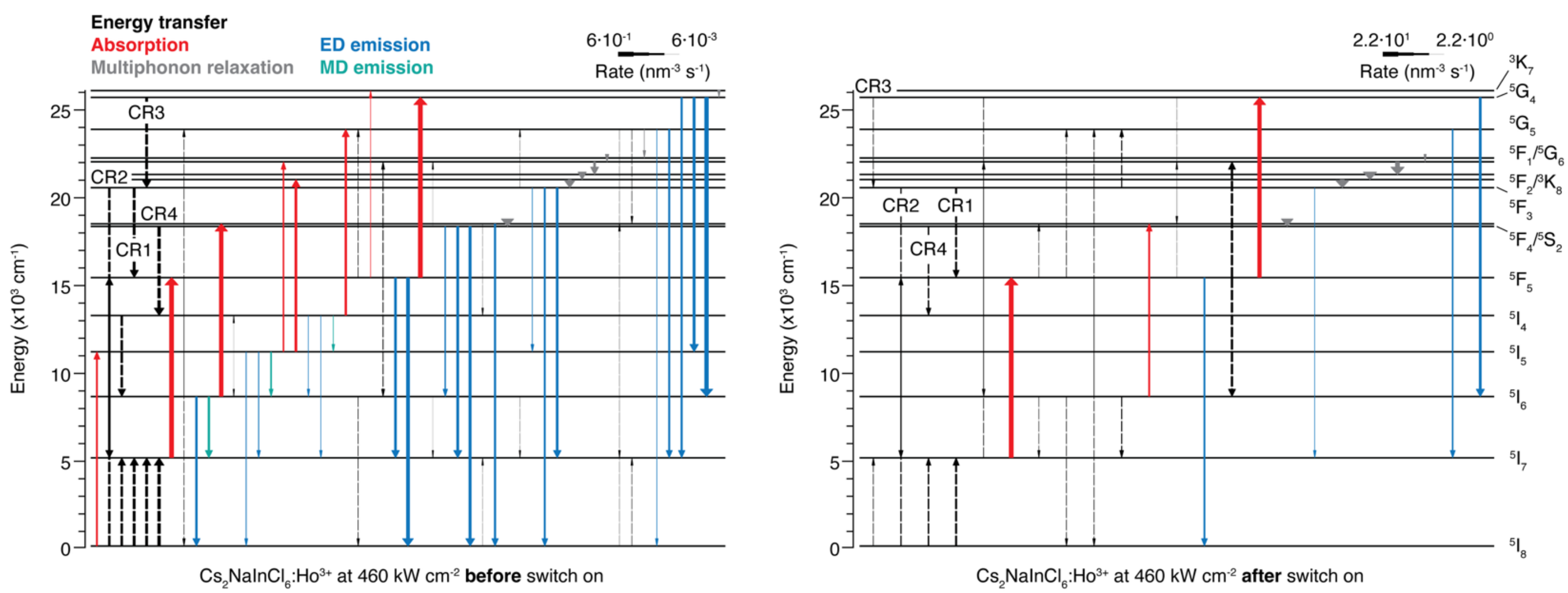

**Supplementary Figure S13.** Simplified energy level diagram and photophysical mechanism in $Cs_2NaInCl_6$: 20% $Ho^{3+}$. The two diagrams represent primary energy transfer (black arrows), photon absorption (red arrows), photon emission (blue and teal arrows), and multiphonon relaxation (gray arrows) processes in $Cs_2NaInCl_6$: 20% $Ho^{3+}$ at the same 976 nm pump power density of 460 kW·cm$^{-2}$ before (forward scan, left panel) and after (reverse scan, right panel) photoswitching. Only transitions with a rate greater than 2.2 nm$^{-3}$·s$^{-1}$ are shown in the right panel. Arrow thickness scales with the transition rate (e.g., thicker arrows represent higher rates).

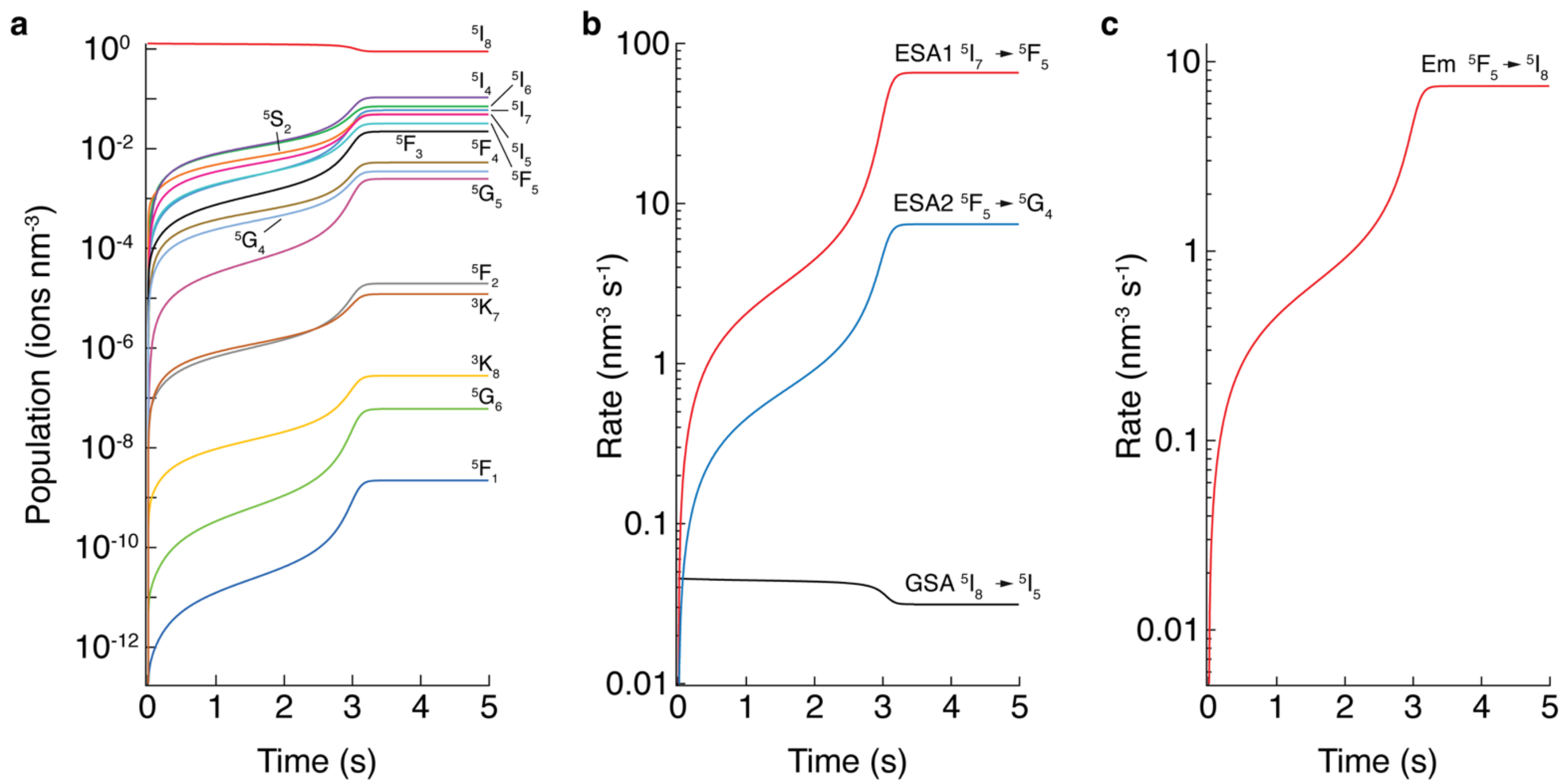

**Supplementary Figure S14.** Time-dependent energy-level population (**a**), absorption (**b**), and emission (**c**) of $Cs_2NaInCl_6$: 20% $Ho^{3+}$ under 976 nm excitation at 500 kW·$cm^{-2}$ starting from a system in its ground state. Note the double-rise profile of the change in excited-state populations and emission, once ESA1 and ESA2 completely take over GSA and significantly depopulate the ground state ($^5I_8$).

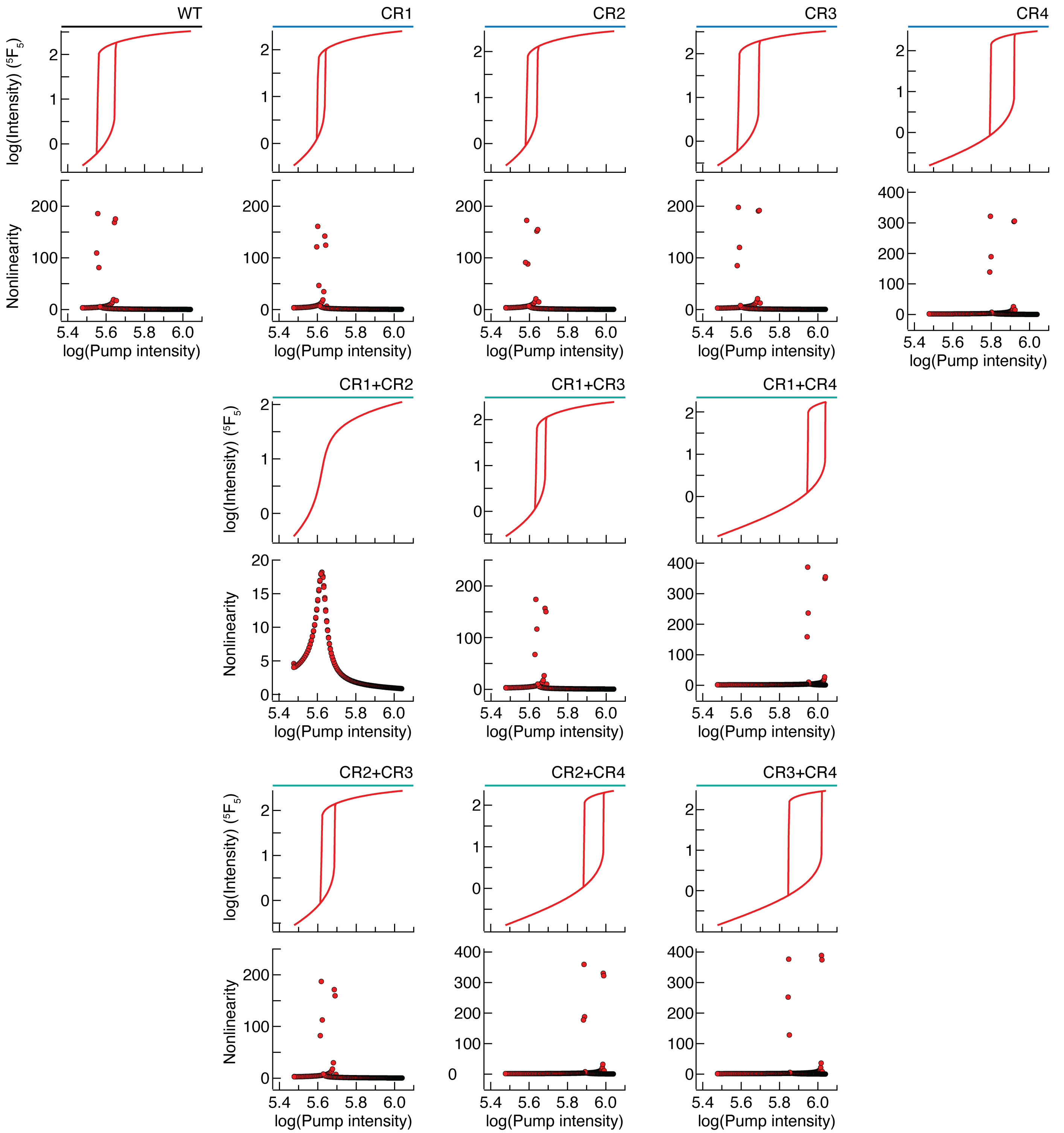

**Supplementary Figure S15.** Simulated pump (976 nm) power dependence of $Ho^{3+}$ emission (660 nm, $^5F_5 \rightarrow {}^5I_8$) in $KPb_2Cl_5$: 15% $Ho^{3+}$. Both forward (increasing power) and reverse (decreasing power) scans are present. Intensity gradient plots (red circles) represent nonlinearity change with power. WT - wild-type simulation, pristine model without knock-outs (KOs); single KOs: CR1 = [$^5F_3$, $^5I_8$] → [$^5F_5$, $^5I_7$], CR2 = [$^5F_3$, $^5I_8$] → [$^5I_7$, $^5F_5$], CR3 = [$^5G_4$, $^5I_8$] → [$^5F_3$, $^5I_7$], and CR4 = [$^5F_4$/$^5S_2$, $^5I_8$] → [$^5I_4$, $^5I_7$]; double KOs: CR1+CR2, CR1+CR3, CR1+CR4, CR2+CR3, CR2+CR4, and CR3+CR4. For complete KO details, refer to **Supplementary Table S5**. Note the absence of luminescence hysteresis in the CR1+CR2 double KO simulation. The knocked-out cross-relaxation processes are also labeled in **Supplementary Figure S16**.

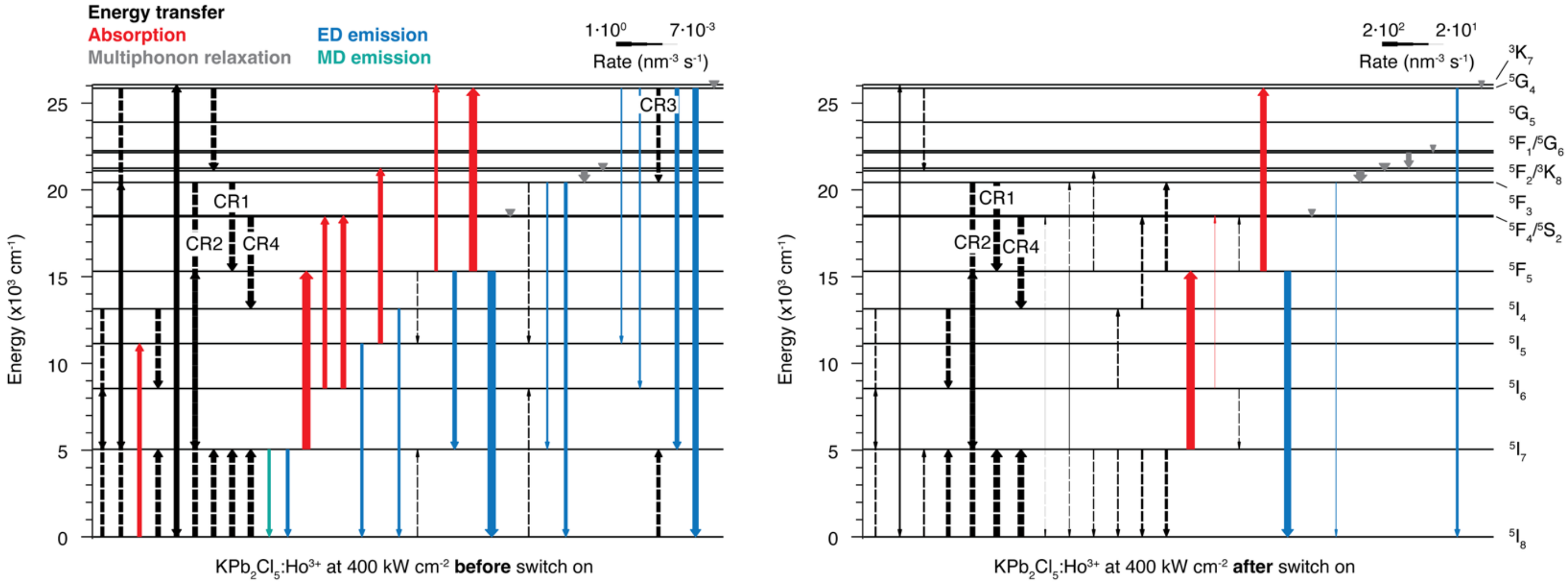

**Supplementary Figure S16.** Simplified energy level diagram and photophysical mechanism in $KPb_2Cl_5$: 15% $Ho^{3+}$. The two diagrams represent primary energy transfer (black arrows), photon absorption (red arrows), photon emission (blue and teal arrows), and multiphonon relaxation (gray arrows) processes in $KPb_2Cl_5$: 15% $Ho^{3+}$ at the same 976 nm pump power density of 400 kW·cm$^{-2}$ before (forward scan, left panel) and after (reverse scan, right panel) photoswitching. Only transitions with a rate greater than 20 nm$^{-3}$·s$^{-1}$ are shown in the right panel. Arrow thickness scales with the transition rate (e.g., thicker arrows represent higher rates).

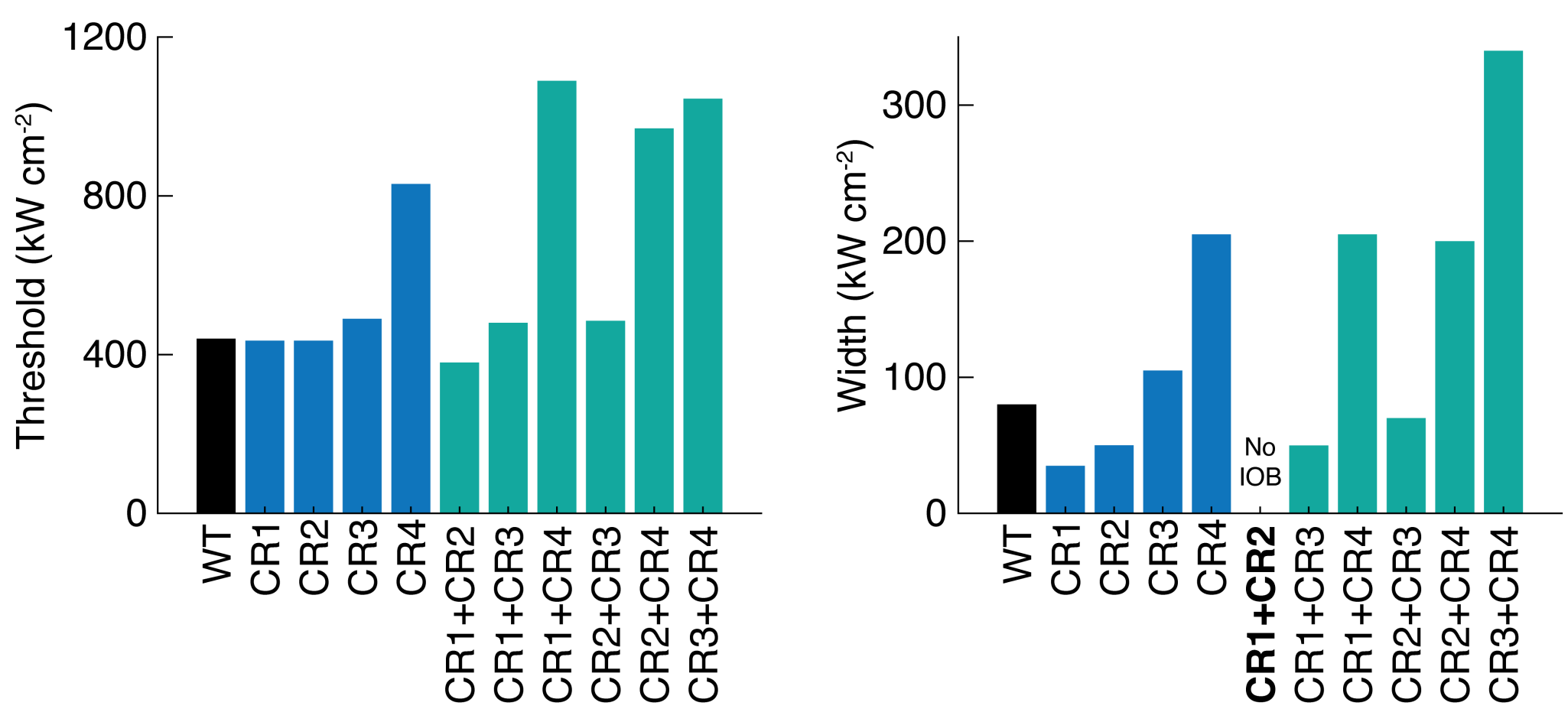

**Supplementary Figure S17.** Influence of CR pathway KOs (shown in Supplementary Figure S15) on the switch-on threshold and hysteresis width in the $KPb_2Cl_5$:$Ho^{3+}$ system, representing $Ho^{3+}$ photoactivation dynamics in a heavy-halide host. No luminescence hysteresis, therefore no IOB, is observed in the CR1+CR2 double KO simulation. Both CR1 and CR2 directly feed the $^5I_7$ and $^5F_5$ reservoirs of $Ho^{3+}$ in XPA. Either cross-relaxation pathway is sufficient to sustain XPA; however, without both, the $Ho^{3+}$ system resembles parallel photon avalanching. Notably, without CR4, the switch-on threshold shifts to higher values, which subsequently widens luminescence hysteresis. We believe that although this pathway is not directly responsible for XPA and IOB, it facilitates seed population at the $^5I_7$ excited state for XPA to activate.

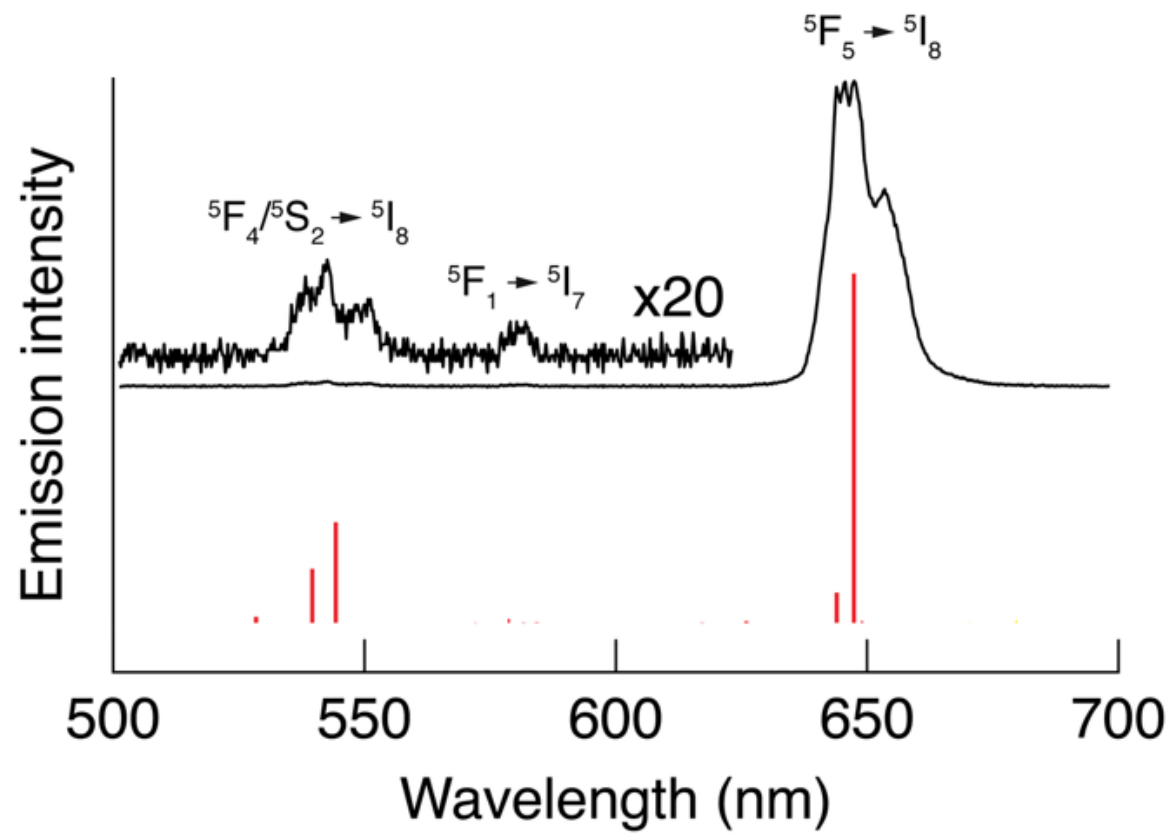

**Supplementary Figure S18**. Experimental (black) and simulated (red) emission spectrum of $NaYF_4$: 20% $Ho^{3+}$. Neither result supports the presence of 590 nm ($^5G_4 \rightarrow {}^5I_6$) emission as in $Cs_2NaInCl_6:Ho^{3+}$. Also see mechanism elaboration in **Supplementary Figure S19**.

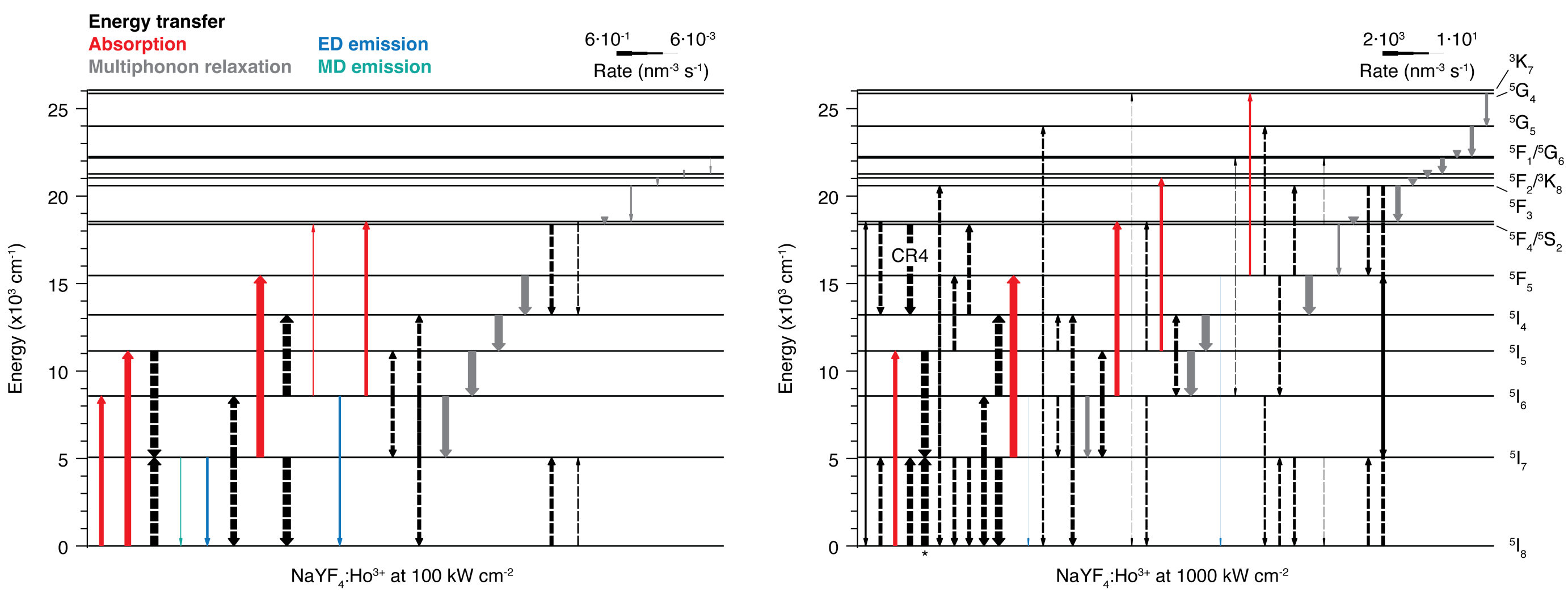

**Supplementary Figure S19.** Simplified energy level diagram and photophysical mechanism in $NaYF_4$: 20% $Ho^{3+}$. The two diagrams represent primary energy transfer (black arrows), photon absorption (red arrows), photon emission (blue and teal arrows), and multiphonon relaxation (gray arrows) processes in $NaYF_4$: 20% $Ho^{3+}$ at the 976 nm pump power density of 100 $kW \cdot cm^{-2}$ (left panel) and 1000 $kW \cdot cm^{-2}$ (right panel). Only transitions with a rate greater than 10 $nm^{-3} \cdot s^{-1}$ are shown in the right panel. CR4 [$^5F_4/^5S_2$, $^5I_8$] → [$^5I_4$, $^5I_7$] and *cross-relaxation [$^5I_3$, $^5I_8$] → [$^5I_7$, $^5I_7$] match previously identified pathways that facilitate parallel photon avalanche upconversion in $NaYF_4:Ho^{3+}$ ANPs as in ref. 22. Arrow thickness scales with the transition rate (e.g., thicker arrows represent higher rates).

Because $NaYF_4$ has relatively large phonon cutoff frequencies, MPR transitions (gray arrows) are prevalent between the $^5I_J$ and $^5F_J$ manifolds, matching the rates of photon absorption and energy transfer.

**Supplementary Section 6. Structural and optical characterization of microcrystalline $Ho^{3+}$-activated double perovskites**

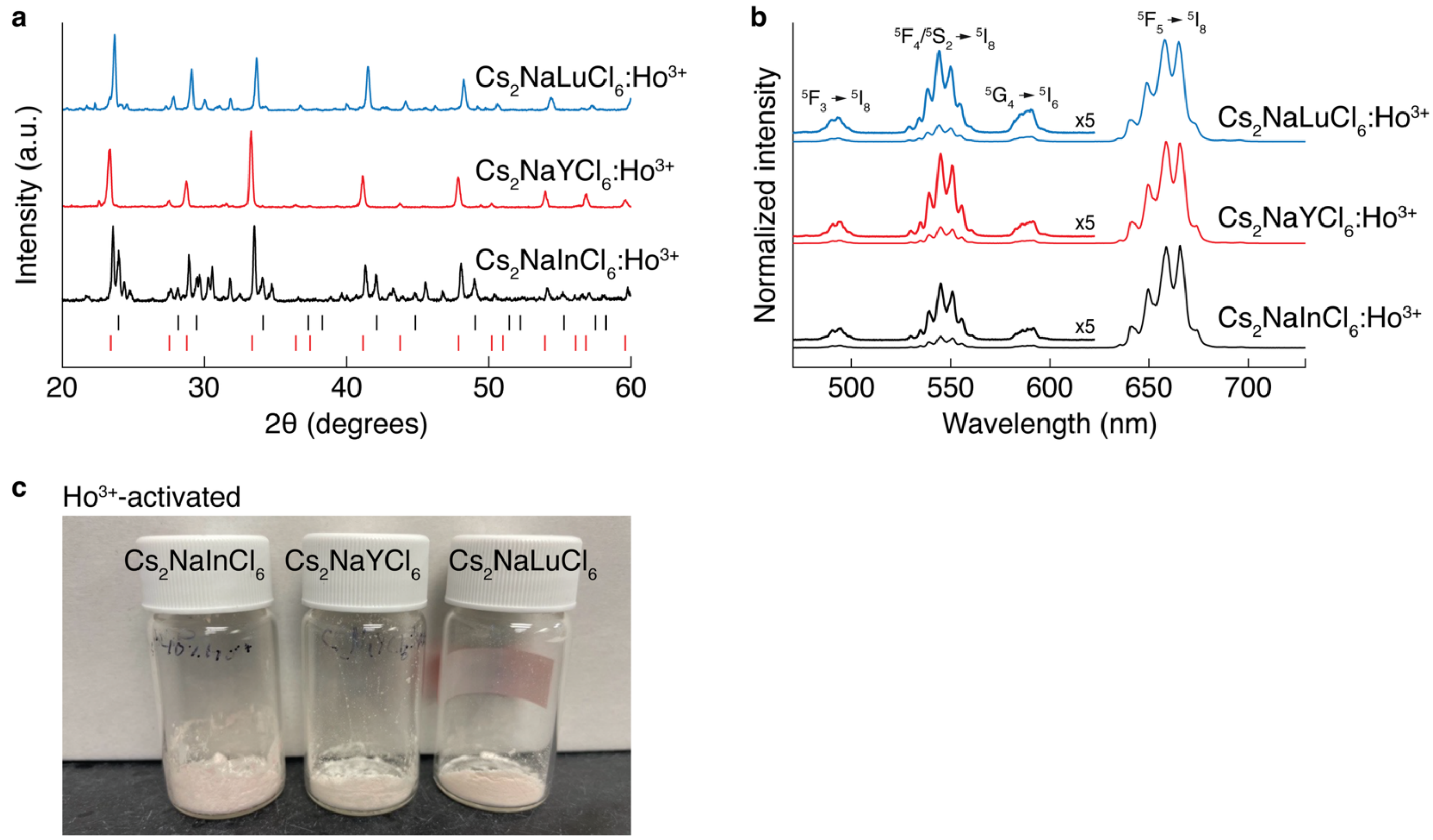

**Supplementary Figure S20. a**, XRD patterns of $Cs_2NaB(III)Cl_6:Ho^{3+}$ (B(III) = In, Y, Lu) microcrystals. Reference patterns for $Cs_2NaInCl_6$ (calculated based on the refined data provided in ref. 6) and $Cs_2NaYCl_6$ (PDF 01-079-0772) are shown as black and red dashes, respectively. The XRD pattern of $Cs_2NaInCl_6:Ho^{3+}$ microcrystals contains the target compound, as well as CsCl, NaCl, and unidentified double-perovskite impurities. **b**, PA upconversion spectra of $Cs_2NaB(III)Cl_6:Ho^{3+}$ (B(III) = In, Y, Lu) microcrystals under 976 nm excitation. Signal intensity in the 475-625 nm range is multiplied by 5-fold for clarity. **c**, Photographs of microcrystalline powders, obtained via dissolution-recrystallization synthesis.

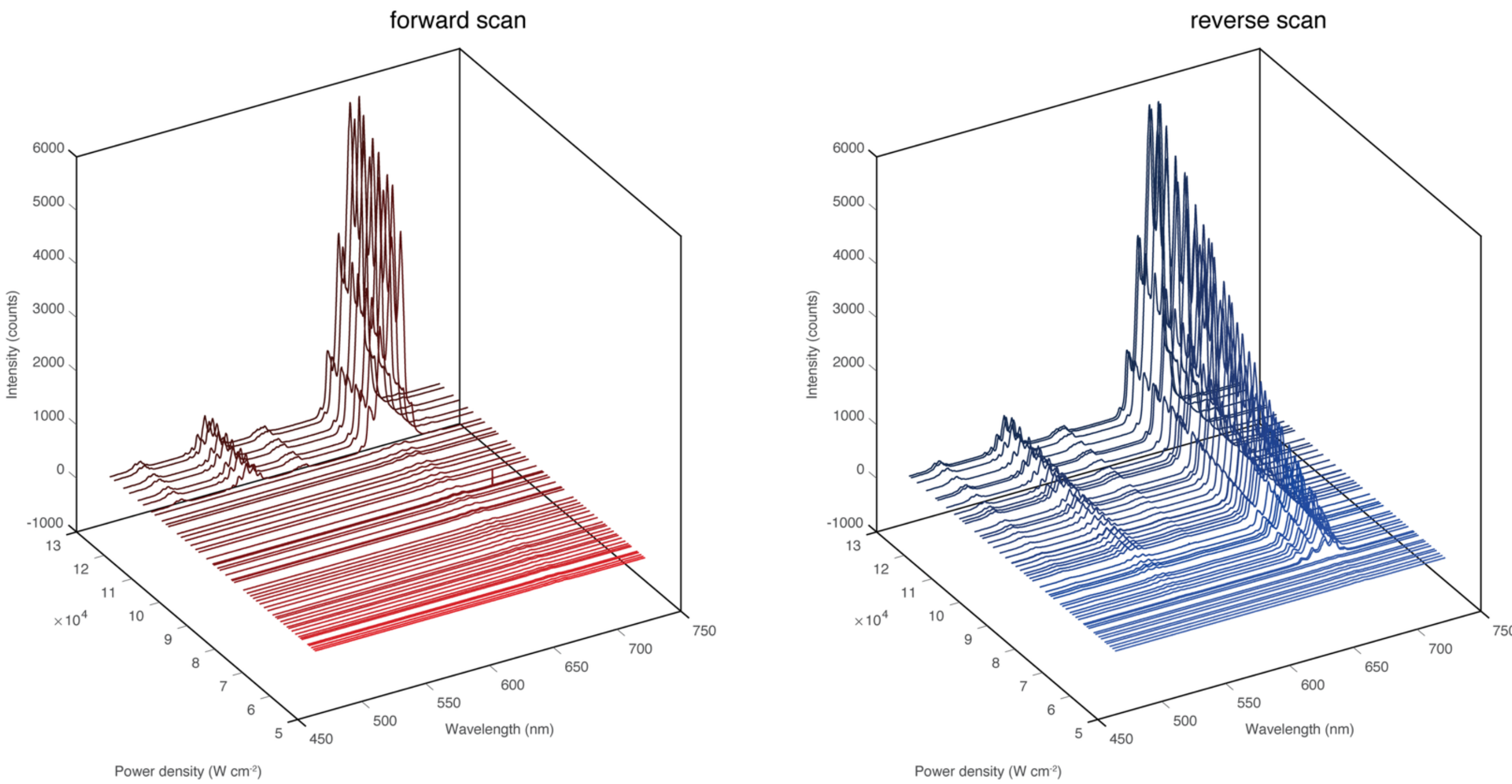

**Supplementary Figure S21.** 976 nm pump power dependence of $Ho^{3+}$ emission spectra in $Cs_2NaInCl_6$: 40% $Ho^{3+}$ microcrystals at room temperature with a spectrometer, instead of an APD as in all other power-dependent measurements. Left – spectra of a forward scan: increasing the pump intensity; right – reverse scan: decreasing the pump intensity. These spectral measurements can be visualized in the video recording of the measurement in **Supplementary Video 1**.

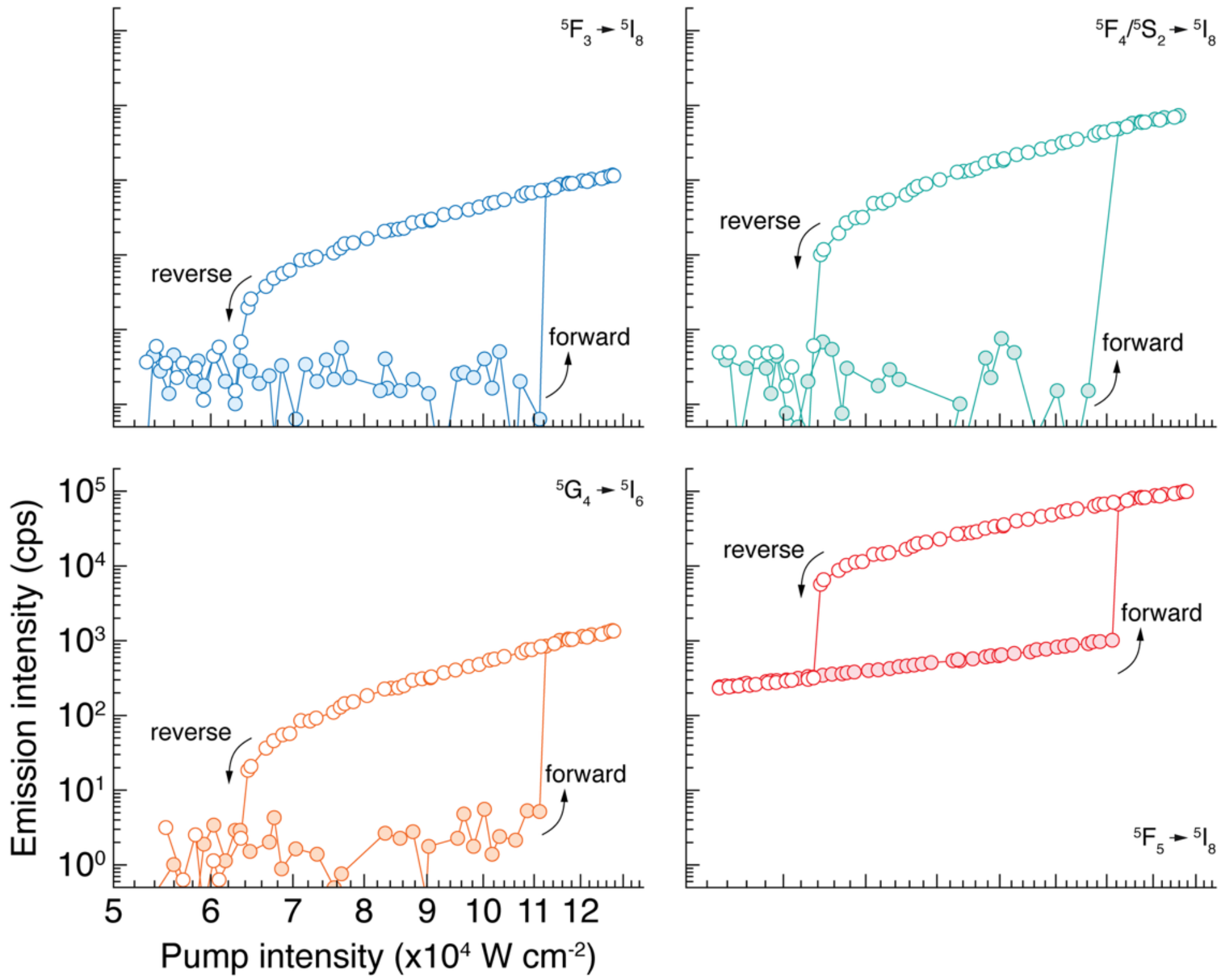

**Supplementary Figure S22.** Integrated $Ho^{3+}$ emission intensity (based on the spectral data in **Supplementary Figure S21**) of $Cs_2NaInCl_6$: 40% $Ho^{3+}$ microcrystals at room temperature against 976 nm pump power. Integrated intensities correspond to $^5F_3 \rightarrow {}^5I_8$ (integration range 484–501 nm), $^5F_4/^5S_2 \rightarrow {}^5I_8$ (integration range 529–562 nm), $^5G_4 \rightarrow {}^5I_6$ (integration range 579–598 nm), and $^5F_5 \rightarrow {}^5I_8$ (integration range 632–675 nm) radiative transitions. Closed circles represent PL intensity acquired at increasing excitation intensity (forward scan), and open circles at decreasing excitation intensity (reverse scan). All graphs share the same x- and y-axis values.

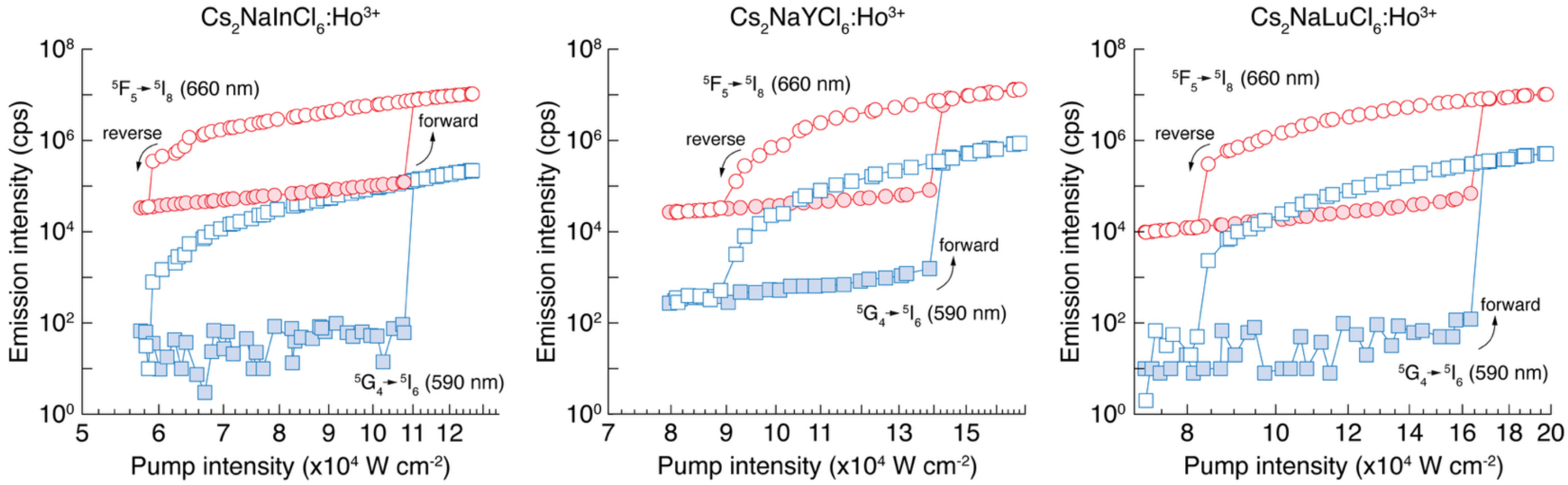

**Supplementary Figure S23.** 976 nm pump power dependence of $Ho^{3+}$ emission at 590 nm ($^5G_4 \rightarrow ^5I_6$; blue squares) and 660 nm ($^5F_5 \rightarrow ^5I_8$; red circles) from $Cs_2NaB(III)Cl_6$: 40% $Ho^{3+}$ microcrystals (B(III) = In, Y, Lu) at room temperature. Filled data points represent the PL intensity measured at increasing excitation intensity (forward scan), and open ones at decreasing excitation intensity (reverse scan).

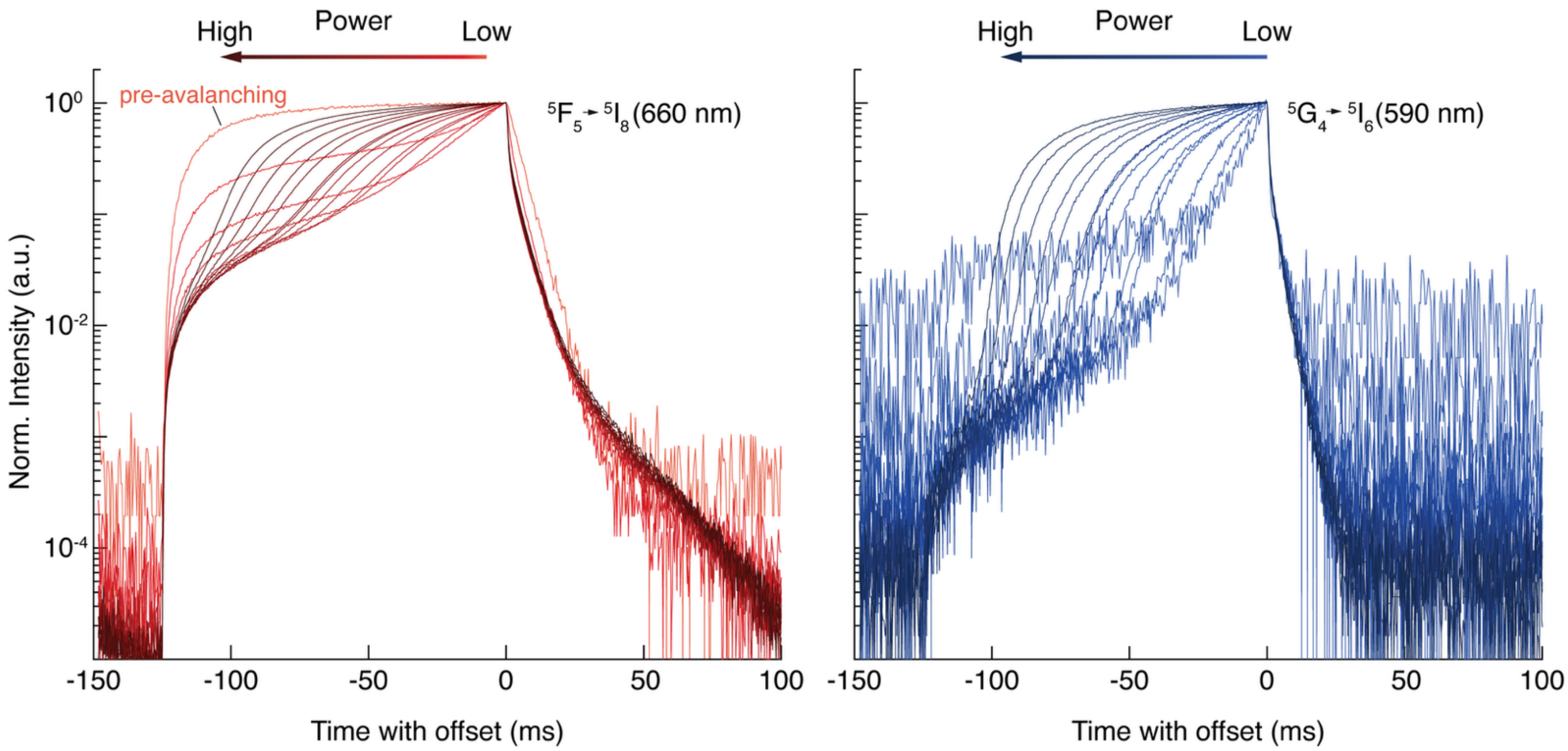

**Supplementary Figure S24**. Time-resolved photoluminescence of $Cs_2NaInCl_6$: 40% $Ho^{3+}$ microcrystals at different 976 nm pump intensities. Right – emission acquired for the 590 nm channel ($^5G_4 \rightarrow {}^5I_6$); left – for the 660 nm channel ($^5F_5 \rightarrow {}^5I_8$). The figure also shows time-resolved luminescence of the 660 nm emission in the pre-avalanching regime. Traces are time-offset to distinguish rise and decay profiles relative to the falling edge of the excitation pulse. The 976 nm pump was pulsed at 4 Hz with a 50% duty cycle.

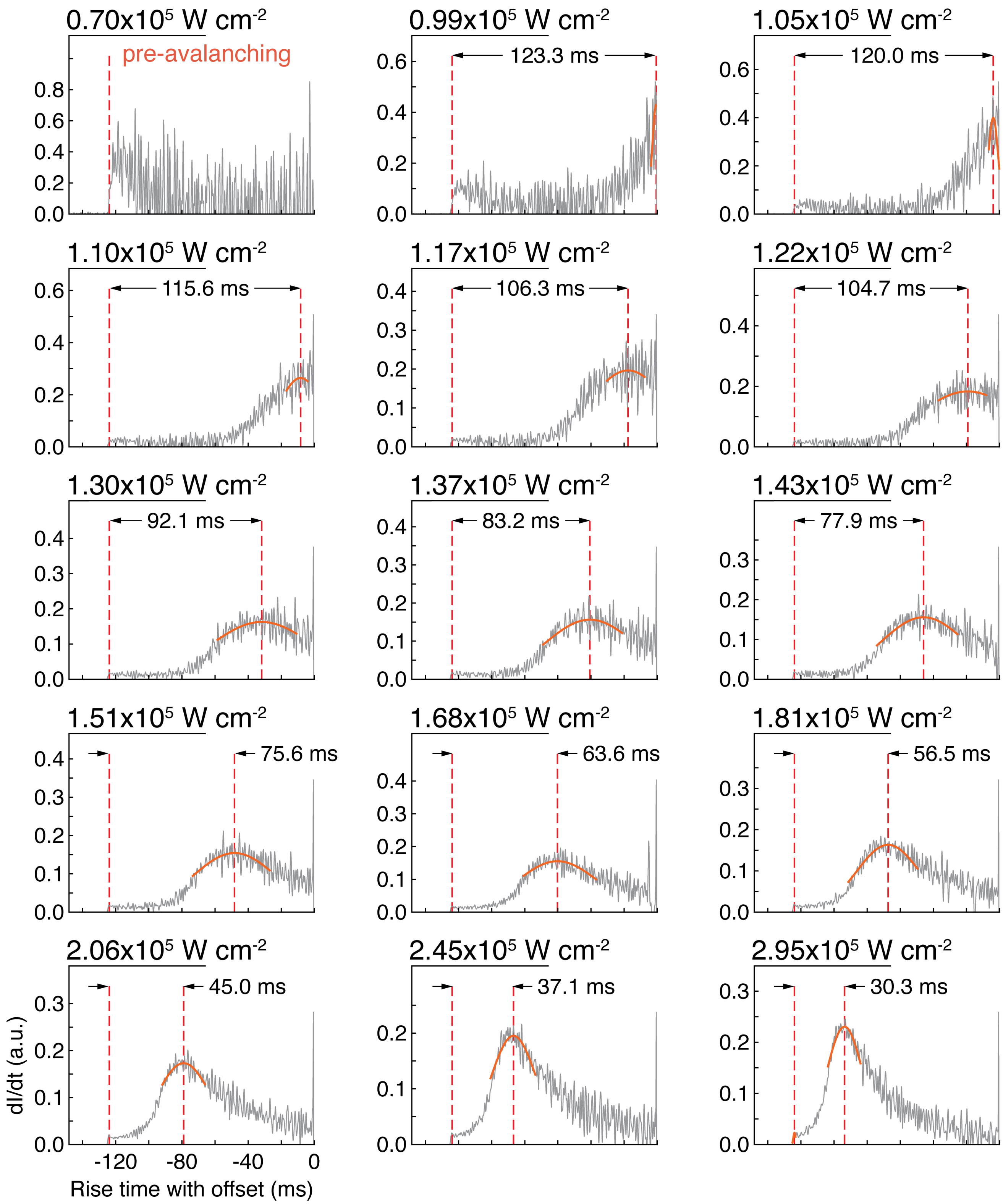

**Supplementary Figure S25**. Derivatives (*dI/dt*) of rise-profile data for 660 nm emission of $Cs_2NaInCl_6$: 40% $Ho^{3+}$ microcrystals, as in **Supplementary Figure S24**. To improve the signal-to-noise ratio, the raw time-resolved data were first binned and smoothed using a Gaussian filter and a Savitzky-Golay filter (polynomial order 3). Peak values of *dI/dt* data were Gaussian-fitted to determine the extrema of the rise profiles. The corresponding average pump power intensities and the delay between the first and second rise-profile extrema are indicated. All graphs share the same x-axis values, offset relative to the falling edge of the excitation pulse.

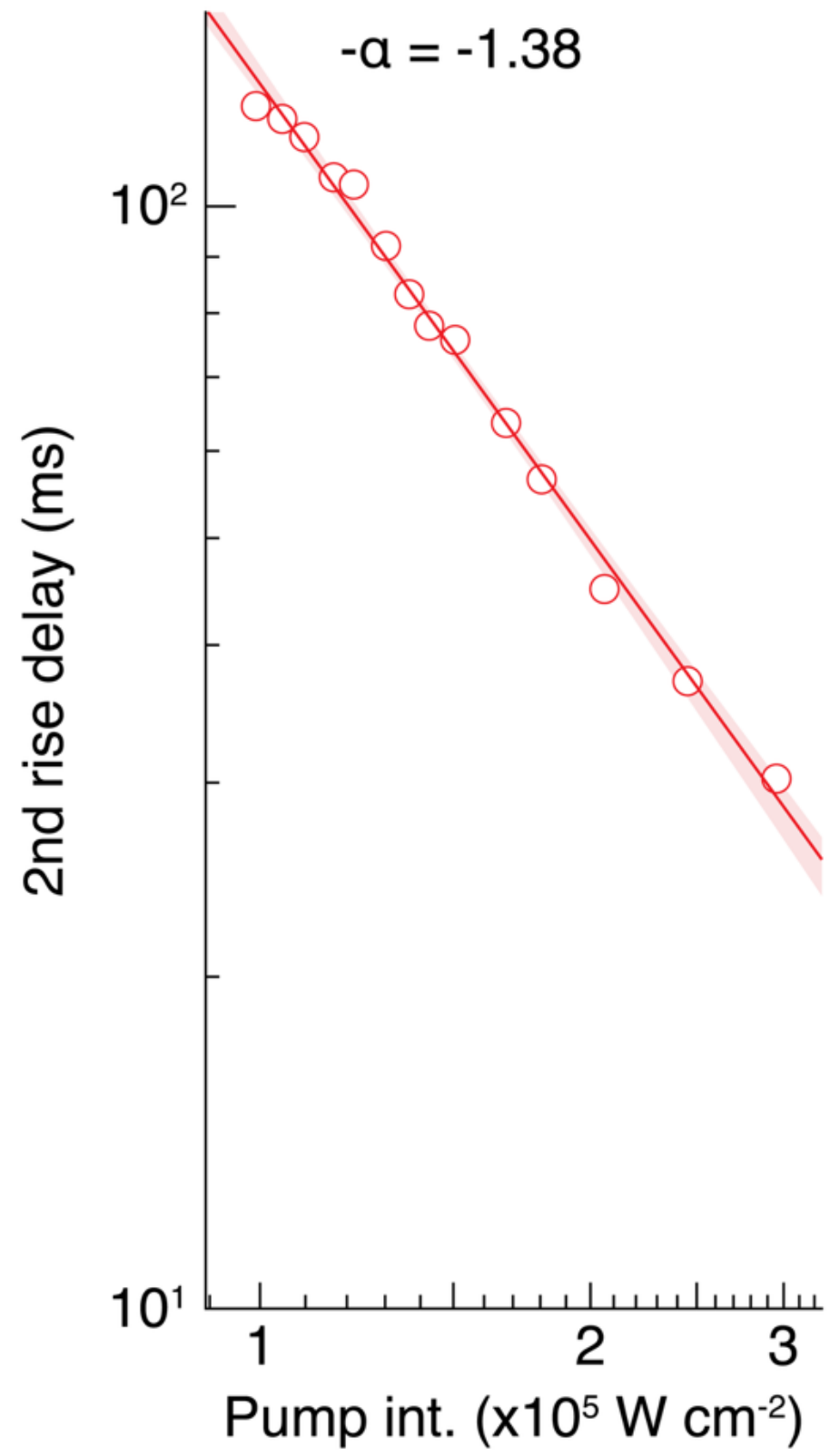

**Supplementary Figure S26**. Log-log plot of the 2nd rise delay vs 976 nm pump intensity (based on **Supplementary Figure S25**). Experimental points (open circles) are fitted to a linear regression (red line), and the 95% confidence interval is indicated by the shaded area. The slope of the fit is $\alpha$.

**Supplementary Section 7. Estimating laser-induced heating of (nano)materials**

Following Baffou et al.,[23] the steady-state temperature increase of a spherical absorber is

$$\delta T = \frac{\sigma_{abs} \cdot I}{4\pi \cdot R \cdot \beta \cdot \kappa_s} \qquad \text{(S21)}$$

$\sigma_{abs}$ is its absorption cross section, $I$ the irradiance, $R$ radius of a spherical absorber, $\beta$ = 1 the dimensionless geometric correction factor of a sphere, and $\kappa_s$ the thermal conductivity of the surrounding medium.

We evaluate two limits for $\kappa_s$: air only (0.025 W·m$^{-1}$·K$^{-1}$) and particles on a glass coverslip in air, $\kappa_s = (\kappa_{air} + \kappa_{glass})/2 \approx$ 0.56 W·m$^{-1}$·K$^{-1}$ with $\kappa_{glass}$ = 1.1 W·m$^{-1}$·K$^{-1}$.

The 976 nm laser excitation was considered at P = 2 mW, greater than that used in power-dependent measurements on $Cs_2NaInCl_6:Ho^{3+}$ (nano)materials, focused by a 0.95 NA objective to a spot of diameter 0.84 μm, giving an irradiance $I$ = 3.6 × 10$^{9}$ W·m$^{-2}$.

We obtained single-ion absorption cross sections $\sigma$ from the simulated rate constants for the $Cs_2NaInCl_6:Ho^{3+}$ system, as 1.4 × 10$^{-26}$ cm$^2$ for GSA ($^5I_8 \rightarrow {}^5I_5$) and 4.5 × 10$^{-22}$ cm$^2$ for the first ESA step ($^5I_7 \rightarrow {}^5F_5$). From the unit-cell volume of $Cs_2NaInCl_6$ (1162.29 Å$^3$, four $In^{3+}$ sites per cell; ref [6]), the $In^{3+}$ density is 3.44 × 10$^{21}$ cm$^{-3}$, and assuming our nominal 40% $Ho^{3+}$ substitution results in $n_{Ho}$ = 1.38 × 10$^{21}$ cm$^{-3}$ dopant density. For a nanoparticle of radius $R$ = 30 nm ($V$ – volume), $\sigma_{abs} = n_{Ho} \cdot \sigma \cdot V$ = 2.2 × 10$^{-25}$ m$^2$ when GSA dominates and 7.0 × 10$^{-21}$ m$^2$ when ESA dominates. The ESA case is an upper limit that assumes every $Ho^{3+}$ ion is in its $^5I_7$ excited state.

A 30-nm-radius nanoparticle in the beam absorbs $Q = \sigma_{abs} \cdot I$ = 7.8 × 10$^{-16}$ W (GSA) or 2.5 × 10$^{-11}$ W (ESA). For $\kappa_s$ = 0.56 W·m$^{-1}$·K$^{-1}$ this gives $\delta T$ = 3.7 × 10$^{-9}$ K (GSA) and 1.2 × 10$^{-4}$ K (ESA). For air only, it results in 8.3 × 10$^{-8}$ K and 2.7 × 10$^{-3}$ K. Extrapolating these values to an ensemble of particles in the beam spot traversing a 1 μm-thick film, the number of irradiated particles with a packing fraction of 1 is around 4900 nanocrystals, resulting in 3.8 × 10$^{-12}$ W (GSA) or 1.2 × 10$^{-7}$ W (ESA) of absorbed power.

Treating the illuminated volume as a sphere of equal volume gives $R_{spot}$ = 0.51 μm and Equation S21 then yields a collective $\delta T$ = 1.1 × 10$^{-6}$ K (GSA) and 3.4 × 10$^{-2}$ K (ESA) on glass, and $\delta T$ = 2.4 × 10$^{-5}$ K (GSA) and 0.8 K (ESA) for air only. Microparticles (≥ 1 μm) exceed the focal spot, so the same collective estimate applies regardless of particle size.

Therefore, assuming all absorbed power converts to heat, sample heating at 2 mW is below 1 K, which is too low to provide significant thermal assistance.

## Supplementary References